\documentclass[longauth]{aa} % for the long lists of affiliations 
\usepackage{amsmath}
\usepackage[table,dvipsnames]{xcolor}% for grey rows on tables
\usepackage{booktabs} %better looking tables with whitespaces
\usepackage{graphicx}

\usepackage{txfonts}
\usepackage[switch]{lineno} % switch permette di posizionare i numeri a sinistra in documenti a due colonne

\usepackage{hyperref}
\hypersetup{
    colorlinks = true,
    linkcolor = blue,
    anchorcolor = blue,
    citecolor = blue,
    filecolor = blue,
    urlcolor = blue
    }

\linenumbers % Attiva numerazione delle righe
\usepackage{array}
\usepackage[flushleft]{threeparttable} % Optional, for footnote alignment
\usepackage{placeins} % reomoves spaces

\newcommand{\orcid}[1]{\href{https://orcid.org/#1}{\includegraphics[width=10pt]{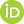}}}

\begin{document} 

   \title{The multi-wavelength evolution of the Luminous Red Nova AT~2021aess}

   \titlerunning{The luminous red nova AT 2021aess}

   \author{F. Guidolin \inst{\ref{uni:merate},\ref{uni:como},\ref{uni:padua}}\orcid{0009-0001-1215-9380}
        \and
        A. Pastorello\inst{\ref{uni:padua}}\orcid{0000-0002-7259-4624}
        \and
        M.~D. Stritzinger\inst{\ref{uni:aarhus}}\orcid{0000-0002-5571-1833}
        \and
        G. Valerin\inst{\ref{uni:padua}}\orcid{0000-0002-3334-4585}
        \and
        A. Reguitti\inst{\ref{uni:padua}}\orcid{0000-0003-4254-2724}
        \and         
        N.~Morrell\inst{\ref{uni:LCO}}\orcid{0000-0003-2535-3091}
        \and 
        J.~P.~Anderson \inst{\ref{uni:santiago}}\orcid{0000-0003-0227-3451}
        \and
        C. R. Burns\inst{\ref{uni:carnegay}}\orcid{0000-0003-4625-6629}
        \and 
        Y.-Z.~Cai \inst{\ref{uni:padua},\ref{uni:YO},\ref{uni:ICS}}\orcid{0000-0002-7714-493X}
        \and
        K. C. Chambers\inst{\ref{uni:hawaii}}\orcid{0000-0001-6965-7789}
        \and
        T.-W. Chen \inst{\ref{uni:GIA}}\orcid{0000-0002-1066-6098}
        \and
        F. Di Mille\inst{\ref{uni:LCO}}\orcid{0000-0003-0483-5083}
        \and
        N.~Elias-Rosa\inst{\ref{uni:padua},\ref{uni:ICE}}\orcid{}
        \and
        M. Fraser\inst{\ref{uni:Dublin}}\orcid{https://orcid.org/0000-0003-2191-1674}
        \and
        L. Galbany\inst{\ref{uni:ICE},\ref{uni:IEEC}}\orcid{0000-0002-1296-6887}
        \and
        J. I. Gonz\'alez Hern\'andez \inst{\ref{uni:Tenerife},\ref{uni:Laguna}}\orcid{0000-0002-0264-7356}
        \and
        M. Gromadzki \inst{\ref{uni:warsaw}}\orcid{0000-0002-1650-1518}
        \and       
        C.~P.~Guti\'errez\inst{\ref{uni:ICE},\ref{uni:IEEC}}\orcid{0000-0003-2375-2064}
        \and           
        W.~B.~Hoogendam\inst{\ref{uni:hawaii}}\orcid{0000-0003-3953-9532}
        \and
        E. Y. Hsiao\inst{\ref{uni:FSU}}\orcid{0000-0003-1039-2928}
        \and
        C.~Inserra \inst{\ref{uni:cardiff university}}
        \and
        E. Kankare \inst{\ref{uni:Turku}}\orcid{0000-0001-8257-3512}
        \and
        T. Kravtsov \inst{\ref{uni:Turku}}\orcid{0000-0003-0955-9102}
        \and
        S. Kumar \inst{\ref{uni:UVA}}\orcid{0000-0001-8367-7591}
        \and
        J. Lu \inst{\ref{uni:MSU}}\orcid{0000-0002-3900-1452}
        \and
        P.~Lundqvist\inst{\ref{uni:stokholm}}
        \and
        K. K. Matilainen\inst{\ref{uni:Turku}}\orcid{0000-0002-8111-4581}
        \and
        P. A. Mazzali\inst{\ref{uni:LJMU},\ref{uni:Max Planck}}
        \and
        T. E. M\"uller-Bravo \inst{\ref{uni:Trinity},\ref{uni:Arturo}}\orcid{0000-0003-3939-7167}
        \and      
        M.~M.~Phillips\inst{\ref{uni:LCO}}\orcid{0000-0003-2734-0796}
        \and
        I.~Salmaso \inst{\ref{uni:Napoli}}\orcid{0000-0003-1450-0869}
        \and        
        M.~Shahbandeh\inst{\ref{uni:JWST}}\orcid{0000-0002-9301-5302}
        \and
        S. Srivastav\inst{\ref{uni:Oxford}}\orcid{0000-0003-4524-6883}
        \and
        S. C. Williams\inst{\ref{uni:Turku}}\orcid{https://orcid.org/0000-0001-8178-0202}
        \and
        D. R. Young \inst{\ref{uni:queen's}}\orcid{0000–0002–1229–2499}
        }
          
\authorrunning{Guidolin et al.}         

    \institute{
    INAF Osservatorio Astronomico di Brera, Via E. Bianchi 46, I-23807 Merate (LC), Italy\label{uni:merate} \label{1}
    \email{francesco.guidolin@inaf.it}
    \and
    Università Insubria, Via Valleggio 11, I-22100 Como (CO), Italy\label{uni:como}\label{2}
    \and
    INAF Osservatorio Astronomico di Padova, Vicolo dell'Osservatorio 5, I-35122 Padova, Italy\label{uni:padua}\label{3}
    \and
    Department of Physics and Astronomy, Aarhus University, Ny Munkegade 120, DK-8000 Aarhus C, Denmark\label{uni:aarhus}\label{4}
    \and
    Carnegie Observatories, Las Campanas Observatory, Casilla 601, La Serena, Chile\label{uni:LCO}\label{5}
    \and
    European Southern Observatory, Alonso de C\'ordova 3107, Casilla 19, Santiago, Chile\label{uni:santiago}\label{6}
    \and
    Observatories of the Carnegie Institution for Science, 813 Santa Barbara St., Pasadena, CA 91101, USA\label{uni:carnegay}\label{7}
    \and
    Yunnan Observatories, Chinese Academy of Sciences, Kunming 650216, P.R. China\label{uni:YO}\label{8}
    \and
    International Centre of Supernovae, Yunnan Key Laboratory, Kunming 650216, P.R. China\label{uni:ICS}\label{9}
    \and
    Institute for Astronomy, University of Hawai'i at Manoa, 2680 Woodlawn Dr., Hawai'i, HI 96822, USA \label{uni:hawaii}\label{10}
    \and
    Graduate Institute of Astronomy, National Central University, 300 Jhongda Road, 32001 Jhongli, Taiwan \label{uni:GIA}\label{11}
    \and
    Institute of Space Sciences (ICE, CSIC), Campus UAB, Carrer de Can Magrans, s/n, E-08193 Barcelona, Spain \label{uni:ICE}\label{12}
    \and
    School of Physics, O’Brien Centre for Science North, University College Dublin, Belfield Dublin 4, Ireland \label{uni:Dublin}\label{13}
    \and
    Institut d'Estudis Espacials de Catalunya (IEEC), Edifici RDIT, Campus UPC, 08860 Castelldefels (Barcelona), Spain \label{uni:IEEC}\label{14}
    \and
    Instituto de Astrof{\'\i}sica de Canarias, E-38205 La Laguna, Tenerife, Spain\label{uni:Tenerife}\label{15}
    \and
    Universidad de La Laguna, Dept. Astrof{\'\i}sica, E-38206 La Laguna, Tenerife, Spain\label{uni:Laguna}\label{16}
    \and
    Astronomical Observatory, University of Warsaw, Al. Ujazdowskie 4,00-478 Warszawa, Poland \label{uni:warsaw}\label{17}
    \and
    Department of Physics, Florida State University, 77 Chieftain Way, Tallahassee, FL, 32306, USA\label{uni:FSU}\label{18}
    \and
    Cardiff Hub for Astrophysics Research and Technology, School of Physics \& Astronomy, Cardiff University, Queens Buildings, The Parade, Cardiff, CF24 3AA, UK \label{uni:cardiff university}\label{19}
    \and
    Department of Physics and Astronomy, University of Turku, 20014 Turku, Finland \label{uni:Turku}\label{20}
    \and
    Department of Astronomy, University of Virginia, 530 McCormick Rd., Charlottesville, VA 22904, USA;\label{uni:UVA}
    \and
    Department of Physics and Astronomy, Michigan State University, East Lansing, MI 48824, USA;\label{uni:MSU}
    \and
    The Oskar Klein Centre, Department of Astronomy, Stockholm University, AlbaNova, SE-10691, Stockholm, Sweden \label{uni:stokholm}\label{21}
    \and
    Astrophysics Research Institute, Liverpool John Moores University, 146 Brownlow Hill, Liverpool L3 5RF, UK\label{uni:LJMU}\label{22}
    \and
    Max Planck Institute for Astrophysics, Karl-Schwarzschild-Str. 1, 85741 Garching, Germany\label{uni:Max Planck}\label{23}
    \and
    School of Physics, Trinity College Dublin, The University of Dublin, Dublin 2, Ireland \label{uni:Trinity}
    \and
    Instituto de Ciencias Exactas y Naturales (ICEN), Universidad Arturo Prat, Chile \label{uni:Arturo}
    \and
    INAF Osservatorio Astronomico di Capodimonte, Salita Moiariello 16, 80131 Napoli, Italy \label{uni:Napoli}
    \and
    Space Telescope Science Institute, 3700 San Martin Drive, Baltimore, MD 21218-2410, USA\label{uni:JWST}
    \and
    Astrophysics sub-Department, Department of Physics, University of Oxford, Keble Road, Oxford, OX1 3RH, UK \label{uni:Oxford}
    \and
    Astrophysics Research Centre, School of Mathematics and Physics, Queen's University Belfast, Belfast BT7 1NN, UK \label{uni:queen's}
        \and
    Institut de C\'{\i}encies de l'Espai (ICE, CSIC), Campus UAB, Cam\'{\i} de Can Magrans S/N, E-08193 Cerdanyola (Barcelona), Spain \label{uni:ICE2}\label{15}
}
   \date{Received 20 10, 2025; accepted 12 11, 2026}

\abstract
    {We present an analysis of the photometric and spectroscopic data of the luminous red nova (LRN) AT 2021aess, in the optical and near-infrared domains. AT 2021aess is the most luminous and one of the longest-lasting LRNe observed to date. The light curve exhibits an initial first peak at $M_V = -15.5 \pm 0.2$ mag, followed by a broader and redder second peak (93d after) at $M_V = -14.2 \pm 0.2$ mag, and a subsequent steep luminosity decline.  
    The extensive photometric monitoring allows us to construct one of the best-sampled LRN light curves, while the spectroscopic observations track the evolution of AT 2021aess from discovery to its second peak. The early spectra display a blue continuum with prominent Balmer emission lines, along with \ion{Fe}{ii} features with P\,Cygni profiles. As the transient evolves, the continuum reddens significantly, the Balmer features weaken, and a forest of metal absorption lines emerges. This spectral evolution is consistent with those previously observed in LRNe.
    Inspecting archival images reveals a progenitor system at the location of the transient, with a total estimated mass in the range $50-70\,M_\odot$.
    These results indicate that AT~2021aess is a key addition to the limited sample of LRNe with massive progenitor candidates, providing new constraints on the most luminous and long-lasting LRNe.}


   \maketitle
   \nolinenumbers

\section{Introduction}

Luminous red novae (LRNe) represent a distinct class of optical stellar transients with luminosities intermediate between those of classical novae and core-collapse supernovae (SNe, see \citealt{Kaminski2026arXiv260517005K} for a recent review). Their peak brightness spans nearly six orders of magnitude, from faint Galactic events such as OGLE 2002-BLG-360 and V1309 Sco (\citealp{Mason_2010}; \citealp{Tylenda_2011,Tylenda_2013}), with absolute magnitudes around M$_V$ $\sim 0$, to luminous extragalactic events reaching M$_V$ $\sim -15$
\citep{Pastorello_2019_jfs}. These ``gap transients'' \citep{Kasliwal_2012, Stritzinger2020a, Pastorello_2019_nature, Cai_2022b} exhibit unique observational signatures, most notably a double-peaked light curve, often interpreted as the result of different stages in the mass ejection and cooling process. In the early stages, their spectra resemble those of Type~IIn SNe, displaying narrow hydrogen emission lines on top of a blue continuum, likely due to interaction with surrounding circumstellar material \citep{Filippenko_1997}. At later times, the spectra become progressively redder and are dominated by molecular absorption bands, such as TiO and VO 
\citep{Pastorello_2019_review,Blagorodnova_2020,Blagorodnova_2021}. 
Compared with intermediate-luminosity red transients (ILRTs), LRNe are generally interpreted as stellar mergers/common-envelope ejection events (e.g., \citealt{Tylenda_2011}), whereas ILRTs are commonly associated with dust-enshrouded super-asymptotic giant
branch (AGB) progenitors and possibly electron-capture SN channels (e.g., \citealt{Botticella_2009, Cai_2021, Stritzinger2020b,Valerin_2025a}). Thus, the two classes are often considered to reflect different dominant evolutionary channels, despite their observational separation is sometimes ambiguous \citep{Kulkarni_2007, Pastorello_2007}.
Although the physical processes leading to LRN outbursts are not yet fully understood \citep{Ivanova_2013c}, these transients are generally interpreted as the outcome of binary interaction leading to common-envelope ejection and, ultimately, stellar coalescence \citep{Tylenda_2011,Ivanova_CEE, Ivanova_2017,Molnar_2017,Pastorello_2023}. This general picture is supported by observations of unstable mass transfer in systems such as V1309 Sco, where Roche-lobe overflow preceded the final coalescence of the stellar components \citep{Mason_2010,Tylenda_2011}. In particular, \citealt{Mason_&_Shore_2022} showed that the ejecta in V1309 Sco are inherently non-spherical and dynamically stratified into distinct high- and low-velocity (HVR/LVR) components, emphasizing how simplified 1D isotropic models are inadequate to map optical depth surfaces and expansion dynamics. In this broad framework, the characteristic double-peaked light curves of LRNe are commonly associated with different phases of mass ejection, cooling, and recombination (e.g. \citealt{Tylenda_2006,Pejcha_2016,Matsumoto_2022}), even if the relative importance of these processes may vary from object to object. 

Observationally, LRNe display a structured photometric evolution. After a long-lasting phase of slow brightening, the light curve shows a sudden luminosity rise to a sharp blue peak, attributed to the ejection of hot material during the merging process. This is followed by a longer-lasting plateau, or even a secondary peak, characterized by much redder colours due to the cooling ejecta \citep{Tylenda_2006, 
Blagorodnova_2017}. From a spectroscopic point of view, only LRN AT~2025abao was monitored during the slow pre-outburst brightening phase \citep{Karambelkar2025,Reguitti_2026b}, showing a continuum consistent with that of a red star. The subsequent spectra, obtained during the first sharp peak, exhibit a blue continuum with prominent Balmer lines in emission, similar to those of other gap transients in which the ejected gas interacts with circumstellar matter (CSM). During the second peak, the spectral continuum becomes increasingly redder, the Balmer emission lines weaken, and numerous narrow metal absorption lines emerge. Finally, in the very late phases, the spectrum shows molecular bands \citep{Martini_1999, Barsukova_2007, 2015A&A...580A..34K, Tylenda_2015}.\\
Despite these general patterns, significant diversity is observed among LRNe, particularly in their photometric evolution. For example, some events lack an early blue peak, which can often be attributed to observational biases (e.g., the first peak may have been missed due to solar conjunction) or to progenitor properties, such as an expanded red giant or supergiant primary star \citep{Pastorello_2021b}. However, the most evident source of diversity among LRNe appears to be related to the behavior of the second light curve peak.
Some events, in fact, display a pronounced red peak, whereas others exhibit a plateau that can last up to 300 days. The duration of this phase and its luminosity greatly vary from one object to another.

Comprehensive datasets capturing all the evolutionary phases of LRNe are essential to advance theoretical models \citep{Pejcha_2016,Segev_2019}. However, the number of well-documented cases, with high cadence, multi-wavelength coverage spanning the early blue peak to the late red decline remains limited, amounting to only around a dozen events to date. A valuable addition to this small sample is provided by the discovery of AT 2021aess, a new member of the class of Gap Transients.
%This paper presents detailed photometric and spectroscopic observations of the LRN: AT 2021aess. 
This object was extensively monitored, and this strategy led to one of the best sampled light curves for a LRN to date, providing valuable insights into the progenitor system and the outburst mechanism.

The structure of this paper is as follows: Sect. \ref{Discovery and host galaxy} contains information regarding the transient and its host galaxy. In Sect. \ref{Photometry}, we present the photometric data and the evolution of the outburst, along with a comparison with a sample of other LRNe and the evolution of their physical parameters (luminosity and color temperature).
Section \ref{Spectroscopic data} describes the spectral evolution of AT 2021aess. The nature of the progenitor is discussed in Sect. \ref{Characterization of the progenitor system}. Lastly, Sect. \ref{Discussion} contains a discussion and the concluding remarks. 

\section{Discovery and host galaxy} \label{Discovery and host galaxy}
\begin{figure}[t]
    \centering
    \includegraphics[scale = 0.36]{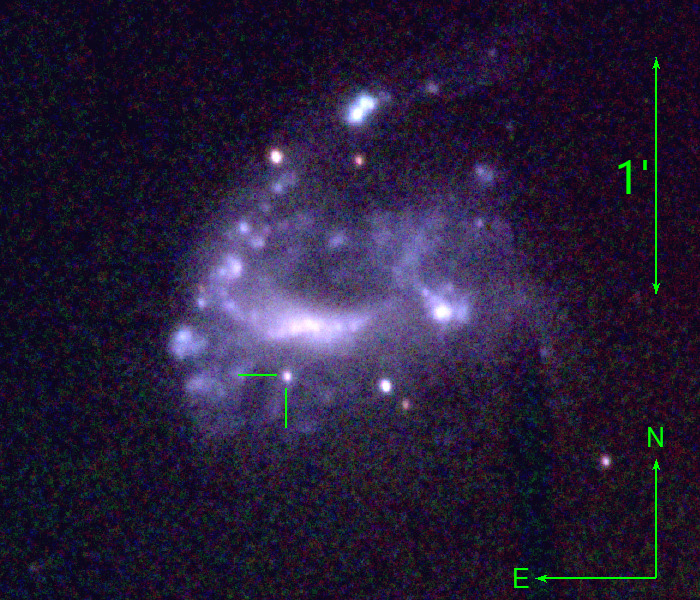}
    \caption{
    Colour image of the host galaxy NGC~1359 and of AT~2021aess taken on 2021 December 15 with the Liverpool Telescope (+IO:O). The image is a combination of the frames obtained with the $B+g$, $V$ and $r$ filters using the RGB technique. The position of the transient is marked.    
    }
    \label{fig:NGC1359}
\end{figure}

AT~2021aess was discovered on November 14, 2021 (MJD = 59532.274) with the 1.2m Samuel Oschin telescope at Palomar Observatory by the Zwicky Transient Facility (ZTF), at the coordinates RA $= 03^{h}33^{m}48\farcs126$, DEC $= -19^\circ29^{m}44\farcs63$ (J2000) \citep{Discovery_astronote}.
At discovery, the new source had a $g$-band AB magnitude of 19.83 mag. AT~2021aess was initially classified on 2021 November 19 as an LBV eruption \citep{Classification_astronote} and later re-classified as a LRN by \cite{Karambelkar_2023}. The host galaxy, NGC 1359, is a nearly face-on irregular galaxy (see Fig.~\ref{fig:NGC1359}).

The distance adopted in this work is \( d = 26.06 \pm 1.83 \) Mpc, corresponding to a distance modulus of \( \mu = 32.08 \pm 0.15 \) mag. This value is based on a kinematic distance, inferred from the redshift of the host galaxy (\( z = 0.006581 \pm 0.000013 \)) using the Hubble law with \( H_0 = 73 \pm 1 \) km s\(^{-1}\) Mpc\(^{-1} \), and assuming a flat cosmology with \( \Omega_m = 0.3 \), \( \Omega_\Lambda = 0.7 \). This choice was preferred over alternative estimates (e.g., Tully-Fisher method) due to the significant scatter and inconsistency in those measurements. For instance, the NASA/IPAC Extragalactic Database (NED) reports markedly different values for the same host, such as \( \mu = 33.28 \pm 0.43 \) mag (\( d = 45.30 \) Mpc) and \( \mu = 31.70 \pm 0.09 \) mag (\( d = 21.90 \) Mpc), which were derived assuming \( H_0 = 75 \) km s\(^{-1}\) Mpc\(^{-1} \) and \( H_0 = 100 \) km s\(^{-1} \) Mpc\(^{-1} \), respectively. The discrepancy among these values reflects the limitations of the Tully-Fisher relation for this specific galaxy, likely due to systematic uncertainties related to inclination, internal extinction, and rotational asymmetries.\footnote{

To refine the adopted distance, we also considered a model-dependent estimate based on \textit{Numerical Action Methods} (NAM), which reconstructs the cosmic orbits of galaxies in the nearby universe by accounting for gravitational influences from surrounding structures, thereby correcting for peculiar velocities (\citealt{Shaya_2017}; \citealt{Kourkchi2020}). We used the publicly available implementation provided by the \textit{Extragalactic Distance Database} (EDD), inputting the supergalactic coordinates of NGC\,1359 (SGL = 259.32°, SGB = --6.48°; \citealt{De_Vaucouleurs_1976}) and its observed heliocentric velocity ($v \approx 1973$\,km\,s$^{-1}$), corresponding to $z = 0.006581$ yielding a distance of $d_{\mathrm{scaled}} = 27.7 \pm 2.1$ Mpc.
This NAM-based estimate closely matches the purely kinematic distance, but has the advantage of correcting for local gravitational influences that dominate at low redshift ($z < 0.01$). 
Independent estimates of the distance to the host galaxy's group environment provide further support for the NAM result. NGC\,1359 is listed as part of a loose group or association centred around RA $\sim 3^{\mathrm{h}}33^{\mathrm{m}}$, Dec $\sim -19.5^{\circ}$ (see \citealt{Kourkchi2020} \& \citealt{Tully2023}, their Group 14-036), with a mean group velocity of $\sim 1950$\,km\,s$^{-1}$ and a group distance ranging from 24 to 27 Mpc. This range is consistent with the NAM prediction and suggests that the galaxy's motion is well captured by local gravitational flow models.}
We also note that NGC\,1359 is not strictly face-on: published axial ratios span (b/a $\approx$ 0.52-0.72; \citealt{LaubertsValentijn1989,deVaucouleurs1991,2MASS}), corresponding to a moderate inclination ($i\sim$ 45$^\circ$-- 60$^\circ$, depending on the adopted isophote and thickness correction). The host is a very late-type galaxy (with a morphologic type code T $\approx$ 9, near the Sm/Im regime; \citealt{LaubertsValentijn1989}), which may further contribute to the scatter of TF-based distances for this object. 
The same host galaxy also hosted the Type~IIb SN~2022yvw, further underscoring its relevance for studies of core-collapse SNe (discovery and classification reported in \citealt{Tonry_2022} and \citealt{Zhai_2022}, respectively).

The Galactic reddening at the coordinates of AT~2021aess is $E(B-V)_{\text{Gal}} = 0.0314 \pm 0.0004$ mag \citep{Schlafly}, and all extinction corrections in this work assume the \citet{Cardelli} extinction law with $R_V = 3.1$.\footnote{Adopting $R_V = 3.1$ is a standard but not unique assumption for the extinction law. As AT~2021aess is hosted in an irregular galaxy, where the dust properties may differ from those observed in the Milky Way, we estimate that the reddening law ranges from $R_V = 2.4$ to $3.4$ as estimated for the Magellanic Clouds (\citealt{Bouchet1985,Gordon2003}).}
To estimate the host-galaxy reddening, we used an empirical relation between the equivalent width (EW) of the interstellar \ion{Na}{i} doublet (\ion{Na}{i}~D) absorption lines and the colour excess $E(B-V)$ \citep[see, e.g.][]{Munari_1997}. The EW of the \ion{Na}{i}~D should be retrieved from the transient spectra corresponding of the very early phases, to avoid contamination from the forest of LRN absorption lines populating later spectra (see Sect. \ref{Spectroscopic data}). In particular, an equivalent width of \ion{Na}{i}~D absorption $EW = 0.85 \pm 0.25\, \AA$ is measured from the spectrum obtained at phase $-6.3$ d from the maximum light (see Sect.~\ref{Spectroscopic data}). Using the empirical relations provided by \citet{Poznanski}, corrected by a factor of 0.86 as pointed out by \cite{Dastidar_2025} and adopting the extinction law of \citet{Cardelli}, we obtain $E(B-V)_{\text{host}} = 0.12 \pm 0.09$ mag. Accounting for the Milky Way reddening towards AT~2021aess ($E(B-V)_{\text{MW}} = 0.03$ according to \citealt{Schlafly} extinction maps) we obtain a total line-of-sight colour excess of $E(B-V)_{\text{tot}} = 0.15 \pm 0.09$ mag.
Given the known dispersion of the \ion{Na}{i}~D versus $E(B-V)$ empirical relations, and lacking other robust constraints (e.g., from  \ion{K}{i} or Diffused Interstellar Bands), the host reddening is treated as an approximate estimate. We also remark that adopting the broader range of reddening values for Magellanic-Cloud-like environments, the absorption would be $A_V = 0.47^{+0.03}_{-0.09} $. 

\section{Photometric data}\label{Photometry}

Our dataset of AT~2021aess includes photometry from the following public surveys: 
ZTF\footnote{Images retrieved through the NASA/IPAC Infrared Science Archive: \url{https://irsa.ipac.caltech.edu/Missions/ztf.html}.}  \citep{Masci_2019, bellm_2019} in the $g$ and $r$ filters \citep{Karambelkar_2023}; the Asteroid Terrestrial-impact Last Alert System\footnote{The template-subtracted forced photometry is released through the Asteroid Terrestrial-impact Last Alert System (ATLAS) data-release interface: \url{https://fallingstar-data.com/forcedphot/queue/}.} \citep[ATLAS;][]{Tonry_2018, Smith_2020} in the $orange$ and $cyan$ filters (hereafter, $o$ and $c$, respectively); and the Panoramic Survey Telescope and Rapid Response
System \citep[Pan-STARRS;][]{Chambers_2016, Flewelling_2020, Magnier_2020} in the $i, z$ and ps1-$w$ (hereafter, $w$) filters.

Furthermore, the photometric follow-up campaign of AT 2021aess made use of numerous instruments and facilities available to the following international scientific collaborations: the Nordic Optical Telescope (NOT) Unbiased Transient Survey 2 (NUTS2),\footnote{\url{https://nuts.sn.ie/}.} the ``advanced" extended Public ESO Spectroscopic Survey of Transient Objects (ePESSTO+),\footnote{\url{https://www.pessto.org/}, \citep[see also][for details on the original PESSTO survey]{smartt_2015}.} and the Precision Observations of Infant Supernova Explosions (POISE).\footnote{\url{https://csp.lco.cl/} see also \cite{Poise}.}

The near-infrared (NIR) dataset consists primarily of observations obtained with the 2.56~m NOT equipped with NOTCam and 3.58~m New Technology Telescope (NTT) of the European Southern Observatory (ESO) equipped with SOFI. Additional early-time data points were obtained from the Keck II + NIRES, P200/TSpec spectra by rescaling the acquisition $K$-band images using the flux calibration of the spectrum.

All images were pre-reduced following standard \texttt{IRAF} routines \citep{Tody_1986, Tody_1993}, including overscan, bias, and flat-field corrections for the optical frames, and the corresponding basic calibrations for the NIR data. Optical and NIR photometry was then performed with the (ECsnoopy; \citealp{SNOOPY})\footnote{ECsnoopy is a package for SN photometry using PSF fitting and/or template subtraction developed by E. Cappellaro. A package description can be found at \url{https://sngroup.oapd.inaf.it/ecsnoopy.html}} package which is specifically designed to handle PSF-fitting photometry of transient sources observed with different instruments and detectors. The software was used for the astrometric registration of the frames, the construction of local PSF models, and the measurement of the transient flux, with template subtraction applied whenever suitable reference images were available.

Given the location of AT 2021aess on the projected background of its host galaxy, particular care was taken to minimise telescope-dependent effects related to the different PSF shapes, pixel scales, and seeing conditions of the various datasets. In all cases, the transient flux was measured through PSF fitting rather than fixed-aperture photometry. For each frame, the PSF was modelled using a set of bright, isolated, and unsaturated field stars located close to the target. The local background was estimated in an annulus centred on the transient position and scaled according to the measured image FWHM, in order to limit contamination from the host-galaxy light and to account, as much as possible, for seeing-dependent background variations. When template subtraction was performed, the reference images were geometrically aligned to the science frames and convolved to match the broader PSF before subtraction, thereby reducing subtraction residuals associated with different seeing conditions.

For the Sloan optical bands, archival Pan-STARRS1 images were used as templates in the filters for which direct archival coverage is available, namely $g, r, i,$ and $z$. For the NIR photometry, archival 2MASS images in $J, H,$ and $Ks$ were used as reference templates, with the 2MASS $Ks$ frame adopted as the closest available template for our $K$ band data. No corresponding archival Pan-STARRS1 templates are available for the $u$ band, nor for the Johnson $BV$. 

In those cases, the transient flux was measured on the original images through PSF fitting with a local background determination, which mitigates host contamination even in the absence of a same-filter template image.

The instrumental magnitudes in the Sloan filters were calibrated using zero point and colour-term corrections inferred from reference stars available in the Pan-STARRS catalogue \citep{Chambers_2016}. Nightly zero-points used to calibrate the instrumental magnitudes from the Swope images were determined relative to a local sequence of stars in the field. This local sequence was calibrated using observations of standard star fields, following the procedure described by \citet{krisciunas2017}. The POISE photometry presented here is reported in the natural photometric system of the Swope telescope; for a detailed description of this system, see \citet{stritzinger2011} and \citet{krisciunas2017}.
The calibration of the Johnson $B$ and $V$ magnitudes was obtained using a catalogue of comparison stars built after converting the Sloan catalogue magnitudes into Johnson magnitudes through the transformation relations of \citet{Chonis_2008}.
The NIR data was calibrated based on the Two Micron All-Sky Survey (2MASS) star catalogue \citep{Skrutskie_2006}.

\subsection{Photometric evolution}
\begin{figure*}
    \centering
    \includegraphics[width = 2.05 \columnwidth]{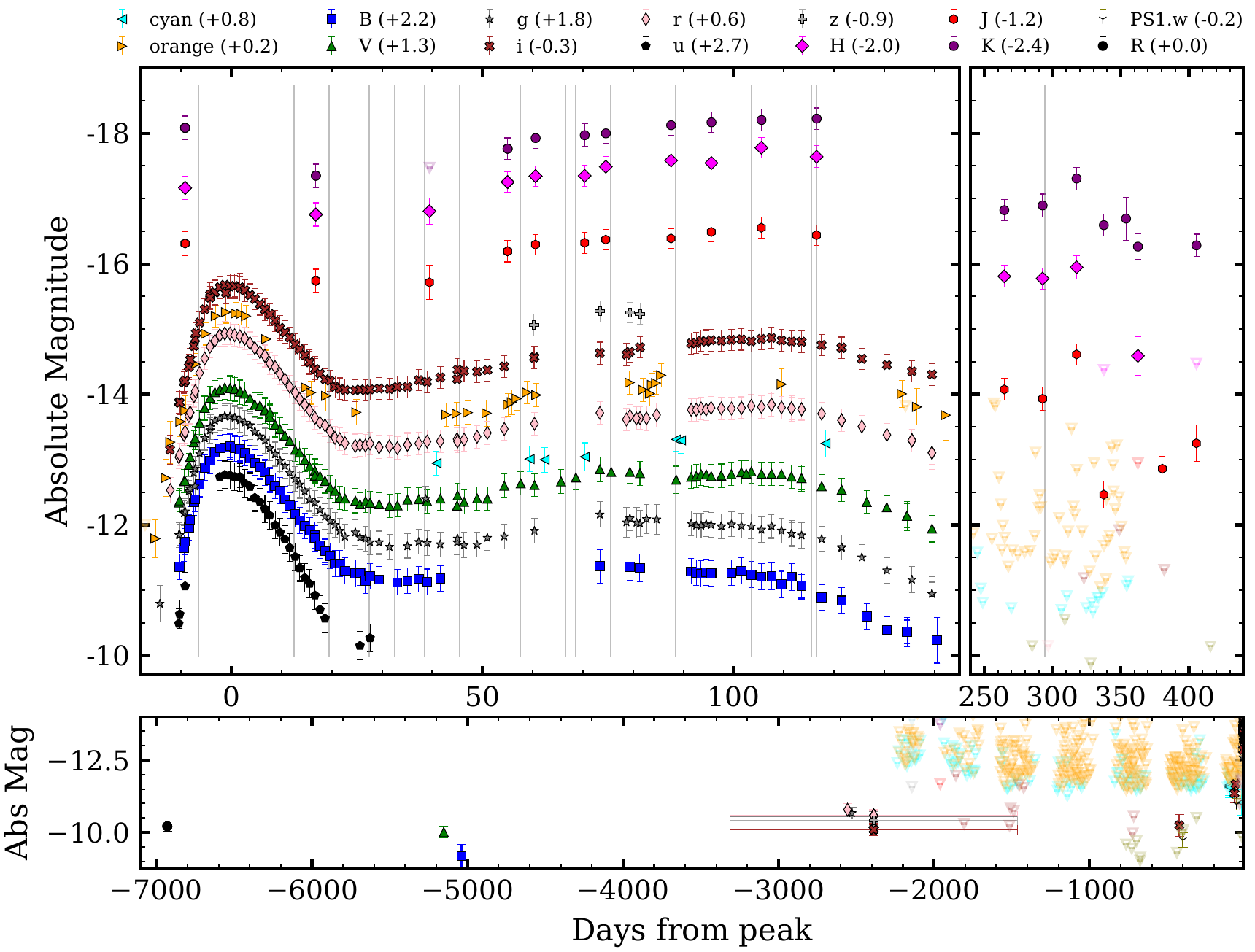}
    \caption{Optical and NIR absolute magnitude light curves of AT~2021aess, with vertical gray lines marking the epochs of spectroscopic observations. The top panels highlight the evolution of the outburst event. There is a gap in observations starting around +150~days after peak brightness, due to the solar conjunction. Observations resumed around phase  +250~days, showing only upper limits in the optical bands, but still multiple detections in the NIR. Magnitudes have been offset for clarity. The bottom panel includes pre-discovery data, in most cases being upper limits, and a limited amount of pre outburst detections.}
    \label{fig:Absolute_magnitude_aess_main_spread_extended}
\end{figure*}

The optical follow-up campaign of AT~2021aess began shortly after its discovery on November 14, 2021 and continued until early January 2023. The photometric evolution of the LRN in the $BVRJHK$ (Vega mag system), $ugriz$ (AB mag) and ATLAS-$c,o$ (AB mag) bands is shown in Fig.~\ref{fig:Absolute_magnitude_aess_main_spread_extended}. The observed magnitudes in all filters, not corrected for line of sight extinction, are reported online.\footnote{A Table is available in electronic form at the CDS, and contains the following information: the epoch and the MJD of the observation (Columns 1 and 2, respectively); the filter (Column 3); the magnitude and the uncertainty (Columns 4 and 5, respectively); the instrumental configuration (Column 6), additional information (Column 7).}

The epochs and magnitudes of the maximum light in the different filters were estimated through 3rd-order polynomial fits (when a sufficient number of data points were available).
The epoch of the $V$-band maximum is used as a reference to compute the phases.
We repeated this procedure to determine the epochs and the magnitudes of the second peak in all filters, which are listed in Table~\ref{mag_peak_2nd} for both optical and NIR bands.

\begin{table*}[htb]
  \centering
  \caption{Epochs and apparent/absolute magnitudes of the 1$^{\mathrm{st}}$ light-curve peak in the different filters, ordered by effective wavelength. \label{mag_peak_1st}}
  \small
    \begin{tabular}{c c c c c c}
    \toprule
    Filter & Date & MJD & Phase (d) & $m$ & $M$ \\    
    \midrule
    $u$ & 2021-11-26 & 59544.54 & $-0.01 \pm 0.50$ & $16.430 \pm 0.020$ & $-15.45 \pm 0.21$ \\
    $B$ & 2021-11-26 & 59544.51 & $-0.04 \pm 0.10$ & $17.285 \pm 0.005$ & $-15.40 \pm 0.20$ \\
    $g$ & 2021-11-26 & 59544.66 & $0.11 \pm 0.35$ & $17.250 \pm 0.005$ & $-15.46 \pm 0.16$ \\
    $V$ & 2021-11-26 & 59544.55 & $0.00 \pm 0.40$ & $17.135 \pm 0.005$ & $-15.39 \pm 0.17$ \\
    $r$ & 2021-11-26 & 59544.96 & $0.41 \pm 0.40$ & $16.805 \pm 0.005$ & $-15.53 \pm 0.17$ \\
    $i$ & 2021-11-26 & 59544.81 & $0.26 \pm 0.25$ & $16.645 \pm 0.005$ & $-15.35 \pm 0.16$ \\
    $o$ & 2021-11-27 & 59545.21 & $0.66 \pm 0.40$ & $16.760 \pm 0.030$ & $-15.69 \pm 0.17$ \\
    \bottomrule
    \end{tabular}
    \tablefoot{The phase represents days sice the first light-curve peak.}
\end{table*}

\begin{table*}[htb]
  \centering
  \caption{Epochs and apparent magnitudes of the 2$^{\mathrm{nd}}$ light-curve peak in the different filters. \label{mag_peak_2nd}}
  \small
    \begin{tabular}{c c c c c c}
    \toprule
    Filter & Date & MJD & Phase (d) & $m$ & $M$ \\    
    \midrule
    $B$ & 2022-02-19 & 59629.12 & $84.57 \pm 1.70$ & $19.150 \pm 0.010$ & $-13.56 \pm 0.20$ \\
    $g$ & 2022-02-20 & 59630.66 & $86.11 \pm 0.97$ & $18.880 \pm 0.005$ & $-13.88 \pm 0.19$ \\
    $V$ & 2022-02-27 & 59637.62 & $93.07 \pm 1.07$ & $18.440 \pm 0.005$ & $-14.11 \pm 0.18$ \\
    $r$ & 2022-03-05 & 59643.73 & $99.18 \pm 0.17$ & $17.925 \pm 0.005$ & $-15.57 \pm 0.17$ \\
    $i$ & 2022-03-07 & 59645.84 & $101.29 \pm 0.43$ & $17.480 \pm 0.005$ & $-14.91 \pm 0.16$ \\
    $c$ & 2022-03-10 & 59648.21 & $103.66 \pm 0.50$ & $18.370 \pm 0.030$ & $-14.16 \pm 0.18$ \\
    $o$ & 2022-03-03 & 59641.27 & $96.72 \pm 2.23$ & $17.810 \pm 0.026$ & $-14.62\pm 0.17$ \\
    $J$ & 2022-03-19 & 59657.55 & $112.90 \pm 1.00$ & $16.800 \pm 0.050$ & $-15.42 \pm 0.16$ \\
    $H$ & 2022-03-18 & 59656.75 & $112.10 \pm 1.00$ & $16.110 \pm 0.050$ & $-16.06 \pm 0.15$ \\
    $K$ & 2022-03-21 & 59659.55 & $114.90 \pm 1.00$ & $16.310 \pm 0.050$ & $-15.82 \pm 0.15$ \\
    \bottomrule
    \end{tabular}
    \tablefoot{The phase represents days sice the first light-curve peak.}
\end{table*}

The light curves of AT~2021aess exhibit a complex evolution that can be broadly described as a sequence of a rapid initial brightening, a swift decline, and a subsequent slower rebrightening. The transient displayed an initial steep rise,  with a rate of $-20.72 \pm 1.53$ mag (100 days)$^{-1}$ in the $V$-band, particularly pronounced in the bluer filters and shallower in the redder bands (see Table~\ref{rates}). This rise culminated in a first (blue) maximum approximately two weeks after discovery, reaching $m_V = 17.135 \pm 0.005$ mag ($M_V = -15.40 \pm 0.18$ mag) at MJD $ = 59544.55 \pm 0.40$ (see Table~\ref{mag_peak_1st}). Immediately following this peak, the light curves underwent a rapid decline across all bands, again more pronounced in the bluer wavelengths, with a decay of $7.5 \pm 0.2$ mag (100 days)$^{-1}$ in the $V$~band. This phase lasted for about 30 days, leading to a relative minimum at $m_V = 18.9 \pm 0.1$ mag ($M_V = -13.60 \pm 0.18$ mag).
Thereafter, a more gradual rebrightening phase ensued, culminating in a broader and fainter second peak at $m_V = 18.4 \pm 0.1$ mag on MJD $ = 59637.62 \pm 0.1$, with a shallower rise rate of $-0.71 \pm 0.06$ mag (100 days)$^{-1}$. This second maximum is only partially covered due to the interruption of follow-up observations during the solar conjunction, which spanned from approximately phase +140 to +250 days. At later times, detections in the NIR filters, starting around $\sim +265$ days, suggest the possible emergence of a third, less pronounced peak, although the data are sparse and scattered.

NIR observations are crucial for LRNe, as the contribution of these bands to the total luminosity of the transient is significant at all epochs, especially during the late phases.
However, due to the limited amount of NIR data during the early phases, it is difficult to obtain a reliable estimate of the rise and decline slopes. The plateau phase is considerably better sampled and, taking $K$ as the reference band, the rise of the second peak presents a slope of $-0.7 \pm 0.09$ mag (100 days)$^{-1}$. After the solar conjunction, as the transient’s luminosity declines significantly, the data are more scattered in time. Nevertheless, there appears to be a third, smaller bump in luminosity at $\sim +320$ days from the first peak.

\subsection{Luminosity and color temperature evolution}\label{Luminosity, radius, and temperature evolution}

Adopting a similar approach as for other LRN studies (see, e.g. \citealt{Cai_2019, Blagorodnova_2020, Blagorodnova_2021, Pastorello_2021b}), we estimate the bolometric light curves and the evolution of the color temperature for AT 2021aess. We emphasize that all derived color temperatures ($T_{\text{BB}}$) are obtained under the assumption of a single-component blackbody radiator.
To compute the SED, we use observations in the Sloan $u, g, r, i$,$o$, Johnson $B$ and $V$, and $J, H, K$ NIR bands, while we exclude the sloan $z$ and $c$ bands because of the limited number of available photometric points. 
When observations in specific filters are missing, we extrapolated the flux value by assuming a constant colour from the previous available epoch. This approach is particularly useful during the initial rise phase, especially for the $u$-band where the sampling is sparse. We divide the interpolation into two time intervals: 

\begin{itemize}
    \item From $-10$ to +35 days: When only the $u,B,g,V,r,o$ and $i$ bands are included due to poor NIR coverage.
    
    \item From +35 to +150 days: When the $u$-band contribution is excluded due to the lack of data, and instead the NIR bands are included in the integration. Given the progressively redder SED at these phases, this approximation has only a modest effect on the integrated flux.

\end{itemize}

\begin{figure}
    \centering
    \includegraphics[width=1\linewidth]{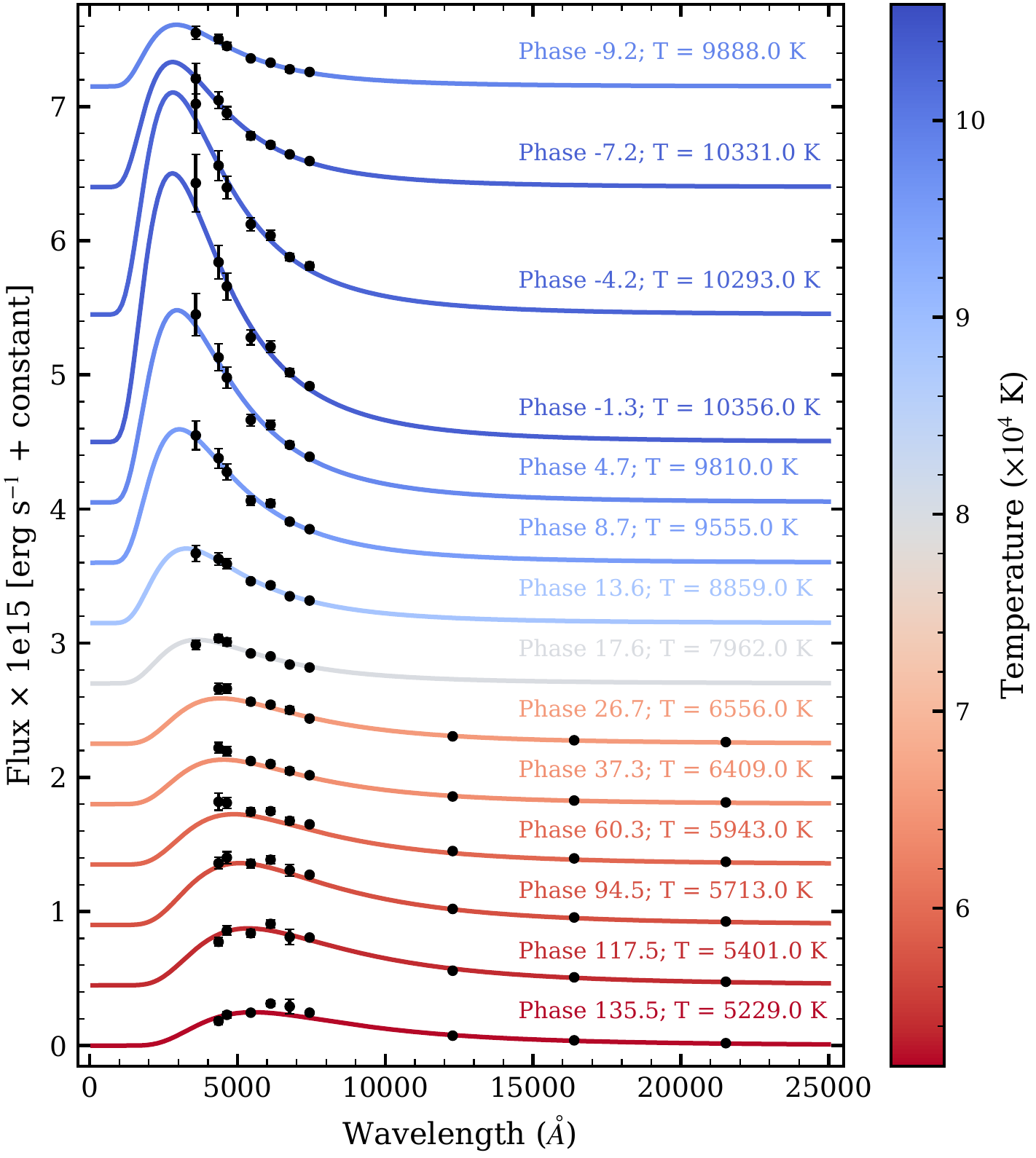}
    \caption{BB fits to the SED of AT 2021aess at different epochs. Fluxes for epochs from +26.7 to +135.5 d have been multiplied by a factor of two for clarity.}
    \label{fig:2021aess_T_L_R_colored}
\end{figure}

\begin{figure}[!h]
    \centering
    \includegraphics[width=1.07\linewidth]{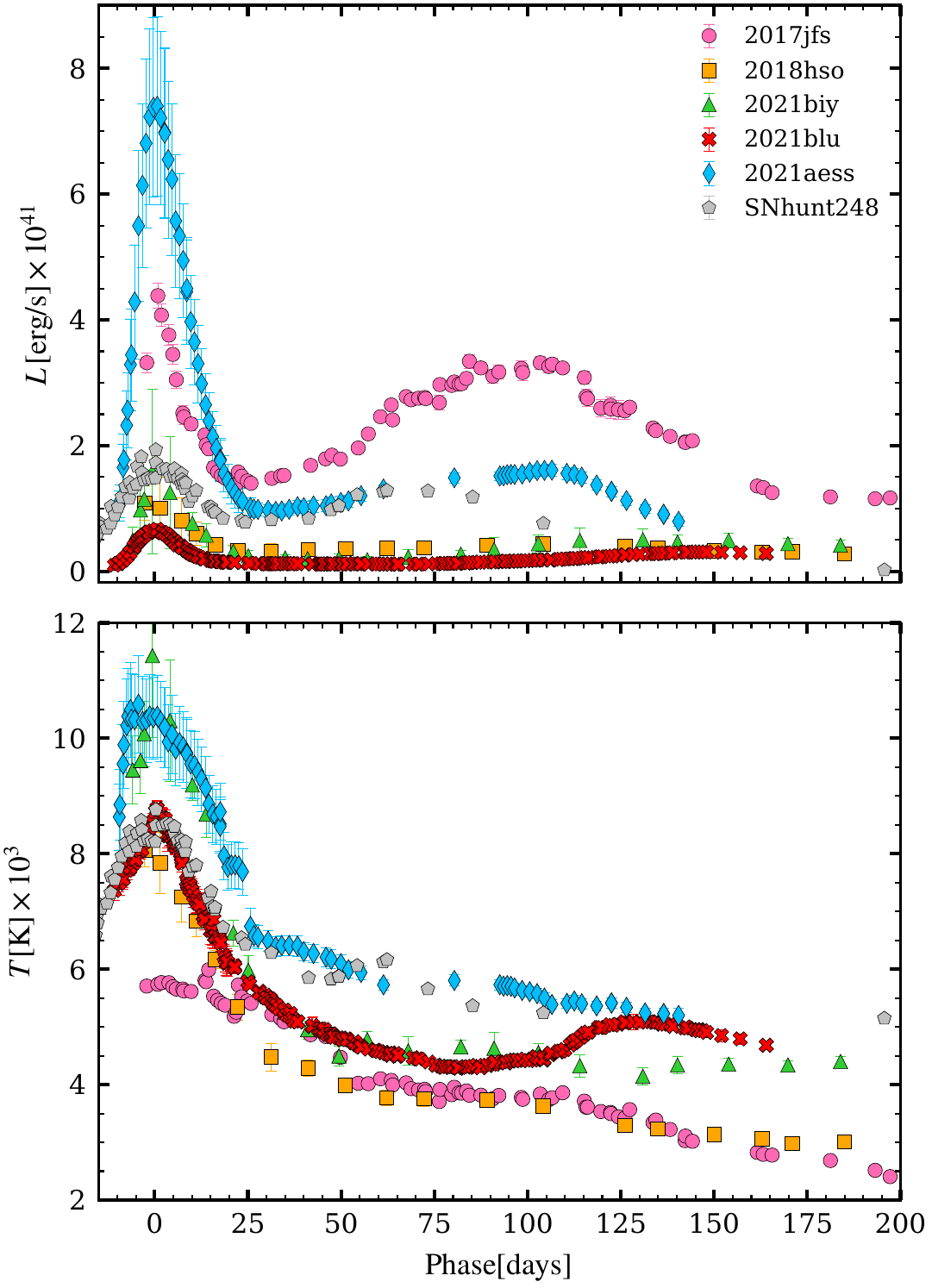}
    \caption{From top to the bottom: bolometric luminosity and color temperature evolution for AT~2021aess, compared to a LRN sample.}
    \label{fig:T_L_R_comparison.PDF}
\end{figure}

At each epoch, we fit the observed broad-band fluxes with a single Planck function to obtain an empirical description of the continuum shape (Fig.~\ref{fig:2021aess_T_L_R_colored}). Although this approximation does not capture the full spectral complexity of LRNe, including line and opacity effects, it provides a reasonable representation of the observed broad-band SED up to the solar-conjunction gap.
 
The resulting bolometric light curve and the blackbody color temperature evolution are shown in the top and bottom panels of Fig.~\ref{fig:T_L_R_comparison.PDF}, respectively, where the bolometric luminosity is obtained by integrating the fitted Planck function over all wavelengths, adopting the distance and reddening discussed in Sect.~\ref{Discovery and host galaxy}.

To constrain the properties of AT~2021aess in the context of LRNe, Fig.~\ref{fig:T_L_R_comparison.PDF} compares its bolometric and color temperature evolution with that of a sample from the literature: SNhunt248 \citep[or AT~2014ib;][]{Kankare_2015, Mauerhan}, AT~2017jfs \citep{Pastorello_2019_jfs}, AT~2018hso \citep{Cai_2019}, AT~2021biy \citep{Cai_2022b}, and AT~2021blu \citep{Pastorello_2023}.
These objects were selected based on similarities in the first-peak luminosity, the plateau duration, and the contrast between the first and second peaks.
AT~2021aess reaches a peak luminosity of $L_{\text{bol}} \sim 7.4 \times 10^{41}$~erg~s$^{-1}$, followed almost four months later by a fainter but broader second maximum with $L_{\text{bol}} \sim 1.7 \times 10^{41}$~erg~s$^{-1}$ (Fig.~\ref{fig:T_L_R_comparison.PDF}, top panel).
Its bolometric output is notably higher than that of the other LRNe, particularly at the first peak, when it is nearly twice as luminous as AT~2017jfs. While all objects reach the post-peak minimum at approximately the same time, the luminosity of AT~2021aess during its $\sim 20-80$~day plateau phase closely resembles that of SNhunt248.

The blackbody color temperature of AT~2021aess peaks at slightly above $10\,300$~K, making it one of the hottest objects in our LRN sample, surpassed only by AT~2021biy at early phases (Fig.~\ref{fig:T_L_R_comparison.PDF}, bottom panel). Following this first maximum, the color temperature drops rapidly to $\sim 6000$~K within 25 days, after which it enters a slower, steady cooling phase, reaching $\sim 5000$~K by $\sim +140$~days. As the transient approaches the second peak, both AT~2021aess and SNhunt248 maintain the highest color temperatures in the sample; only at phase $\sim +130$~days, when AT~2021blu reaches its own second maximum, its color temperature becomes comparable to that of AT~2021aess.

\section{Spectroscopic data}\label{Spectroscopic data}

Our spectroscopic monitoring campaign of AT~2021aess spans four months, from November 19, 2021 (phase $-6.3$ days from the first peak) to March 22, 2022 (phase $+115.5$ days), during which we collected 18 optical (including one of the host galaxy) and 3 NIR spectra. These spectra were obtained in the framework of three major international collaborations: the ePESSTO+\footnote{\url{https://www.pessto.org/index.py?pageId=61}} provided the ESO 3.58-m NTT+EFOSC2 spectra, NUTS2 provided the 2.56-m NOT+ALFOSC spectra, POISE provided all the 6.5-m Magellan (IMACS and FIRE) spectra and various other facilities were made use of including the 10-m Keck+NIRES, and the 2.4-m Hiltner telescope. 
The basic information on the spectra presented in this work is reported in Table~\ref{spectra_log}. The spectral resolutions reported in the table are measured from the full width at half maximum (FWHM) of the [\ion{O}{i}] $\lambda$5577 night-sky line. 
\begin{figure*}\centering
    \includegraphics[width=1.0\linewidth]{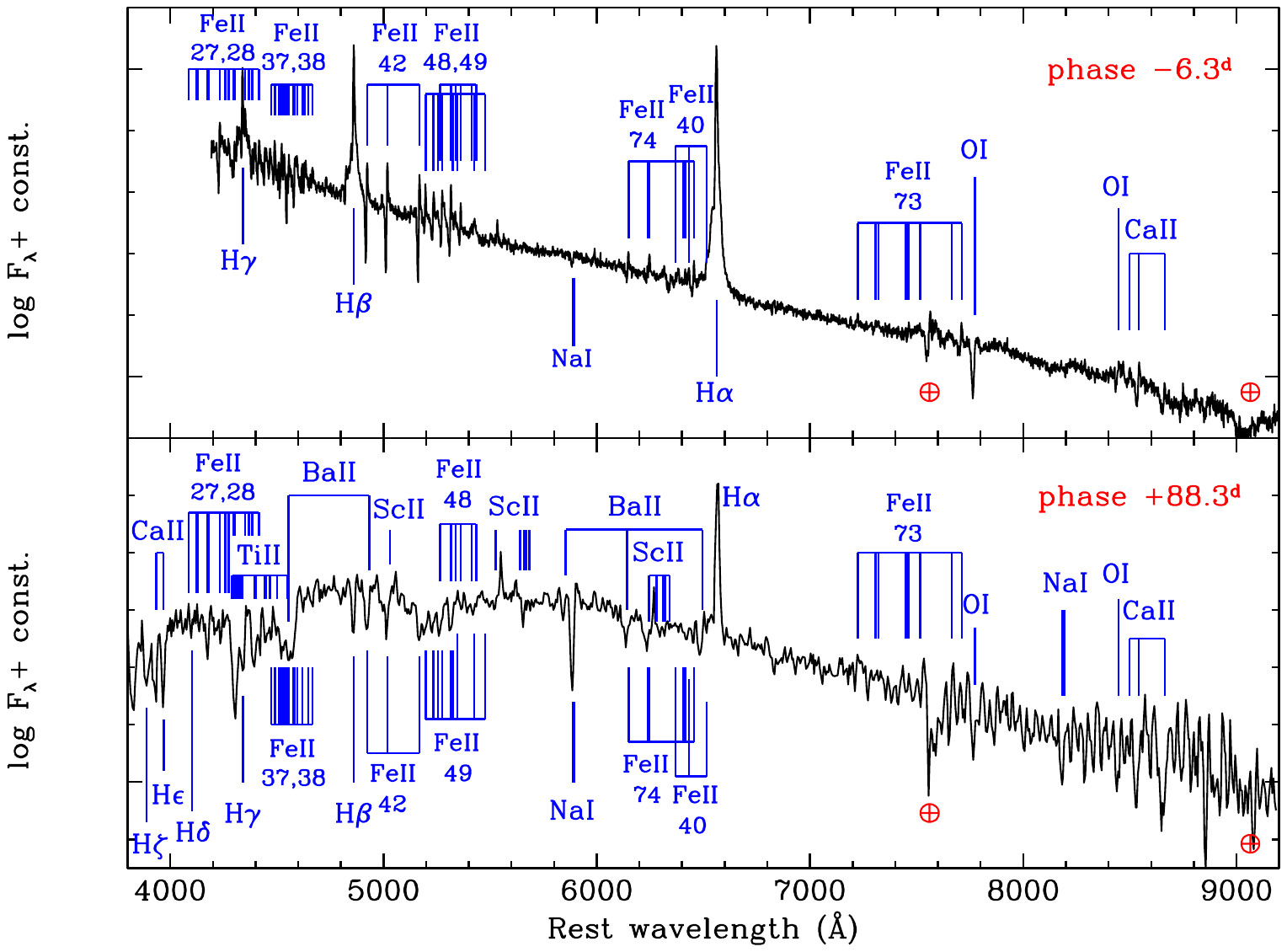}
    \caption{Identification of spectral features in two representative spectra of AT~2021aess. Top panel: Early spectrum (phase $-$6.3 days taken with Baade+IMACS), characterized by a blue continuum and prominent H emission lines. Bottom panel: spectrum of AT~2021aess during the second peak (phase $+$88.3 days taken with NTT+EFOSC2). The spectral continuum, shifted to a redder colour, is dominated by a dense array of metal absorption lines. }
    \label{fig:lineid_rotated.pdf}
\end{figure*}
On September 16, 2022 (+295 days), a spectrum of the location hosting the transient was obtained with the 2.54-m Isaac Newton Telescope (INT) with the Intermediate Dispersion Spectrograph (IDS). The spectrum, shown in Fig.~\ref{fig:galaxy}, is useful to constrain the oxygen abundance at the LRN location (see Appendix~\ref{Host galaxy metallicity estimation}).
The spectra provided by the NUTS2 collaboration were reduced using the \texttt{FOSCGUI}\footnote{FOSCGUI is a graphical user interface designed for extracting supernova spectroscopy and photometry from FOSC-like instruments. Developed by E. Cappellaro, further information can be found at: \texttt{sngroup.oapd.inaf.it/foscgui.html}.} pipeline \citep{Mattila_2016}; 
the spectra from the NTT telescope were reduced with a \texttt{pyraf}-based pipeline (\texttt{PESSTO}\footnote{https://github.com/svalenti/pessto}), optimised for the EFOSC2 instrument,  while data from other instruments were processed using standard routines within the \texttt{IRAF} environment. 
For all datasets, the reduction steps included bias subtraction, flat-field correction, removal of background and cosmic rays, optimal 1D spectral extraction, wavelength and flux calibration, and telluric correction. The resulting spectral flux calibration was cross-checked against contemporaneous photometric measurements, and when discrepancies were identified, a constant scaling factor was applied to ensure consistency. 
In our analysis, the spectra are rest frame corrected adopting a redshift of $z = 0.006581 \pm 0.000013$ and are corrected for a total colour excess towards AT~2021aess of  $E(B-V)_{\text{tot}} = 0.15 \pm 0.09$ mag (see Sect.~\ref{Discovery and host galaxy}).

\subsection{Line identification}

To determine the chemical composition of the material ejected by AT~2021aess, we perform line identification using good resolution and signal-to-noise (S/N) spectra obtained in two representative epochs of the evolution of the LRN: during the rise to the first blue peak (November 19, 2021 phase $-6.3$~days) and at the time of the second red peak (February 22, 2022 phase $+88.3$~days). The lines identified in the two spectra are shown in Fig. \ref{fig:lineid_rotated.pdf}. 

The early optical spectrum of AT~2021aess (Fig.~\ref{fig:lineid_rotated.pdf}, top panel) is characterised by a blue continuum, with dominant H emission lines, prominent \ion{Fe}{ii} features from well-known multiplets (27, 28, 37, 38, 40, 42, 48, 49, 73, 74, e.g. multiplet~42 at $\lambda\lambda$ 4923.9, 5018.4, 5169.0~\AA), along with weaker \ion{Na}{i}~D (D$_2$ 5889.95~\AA, D$_1$ 5895.92~\AA), \ion{O}{i} (e.g. $\lambda$ 8446 and the triplet $\lambda\lambda$ 7771.9, 7774.2, 7775.4~\AA) and \ion{Ca}{ii} lines (including H\&K $\lambda\lambda$ 3933.7, 3968.5~\AA\ and the NIR triplet $\lambda\lambda$ 8498.0, 8542.1, 8662.1~\AA), mostly in absorption. At this epoch, the \ion{Na}{i}~D absorption feature is narrow and unresolved, and is therefore most likely dominated by interstellar gas in the host galaxy. However, given the available spectral resolution, a minor contribution from the transient cannot be securely excluded. More generally, the simultaneous presence of \ion{Na}{i} and \ion{Fe}{II} does not require uniform physical conditions, since these lines may arise in different regions of a spatially extended, density-stratified, and thermally inhomogeneous outflow. In this context, any transient \ion{Na}{i} contribution would plausibly originate in cooler outer layers, while \ion{Fe}{II} would trace deeper and/or hotter regions. We note that, regardless of whether the weak \ion{Na}{i}~D absorption in the earliest spectrum is purely interstellar or includes a small circumstellar material contribution, the overall interpretation remains unchanged. At later phases, the \ion{Na}{i}~D profile becomes stronger and also develops an emission component, indicating a clearer contribution from the transient ejecta.

\begin{figure*}
    \centering
    \includegraphics[width=0.85\linewidth]{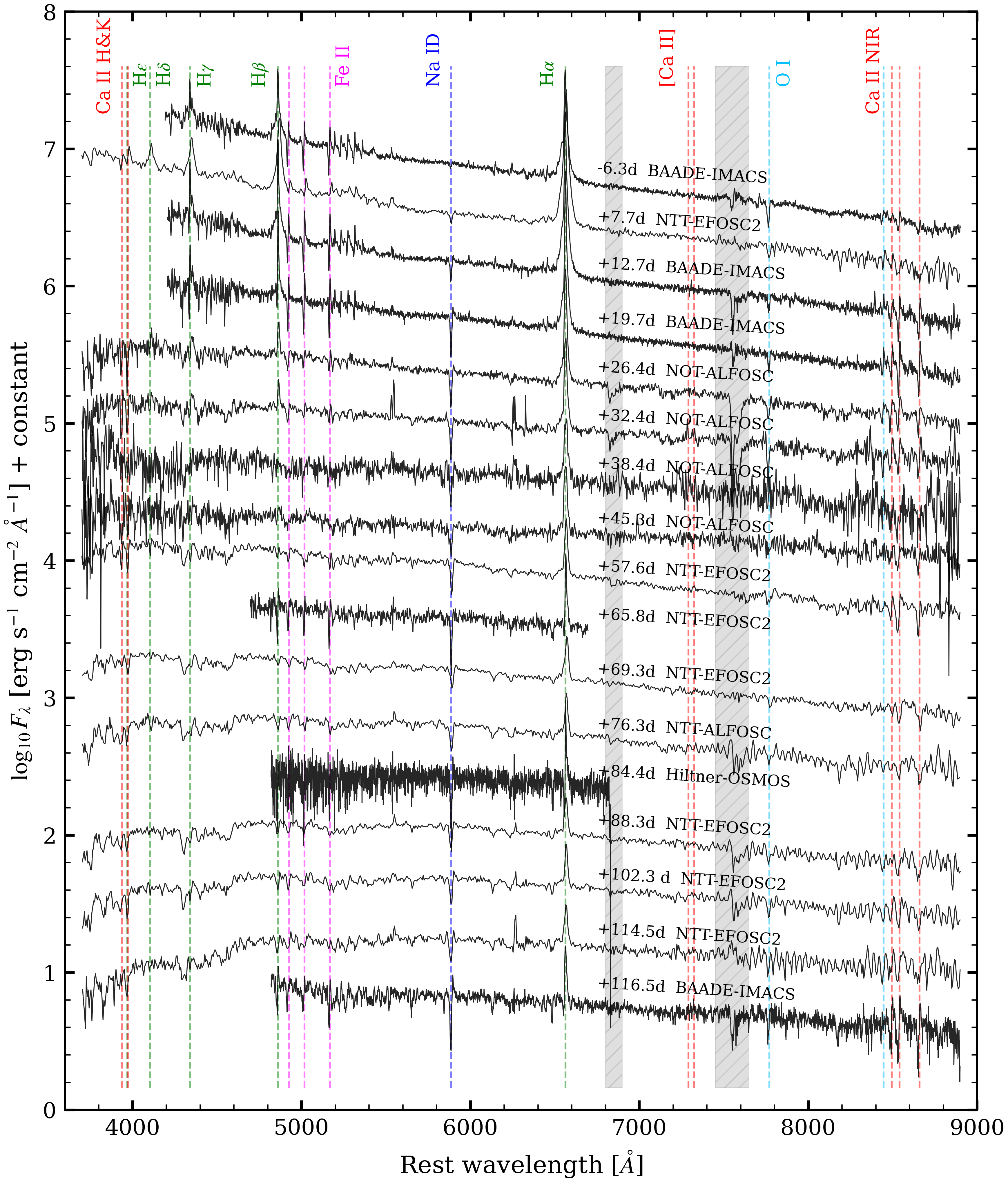}
    \caption{Sequence of optical spectra of AT~2021aess. The most significant spectral features are marked with vertical lines. The grey shaded area highlight the positions of the strongest telluric absorption bands. 
    \label{fig:spectra_evol}}
\end{figure*}

In the optical spectrum at phase $+88.3$~days (Fig.~\ref{fig:lineid_rotated.pdf}, bottom panel), we identify the lines of the Balmer series from H$\alpha$ (6562.8~\AA) to H$\zeta$ (3889.1~\AA), with H$\epsilon$ (3970.1~\AA) partially blended with the \ion{Ca}{ii} H absorption feature. We note that only H$\alpha$ maintains a dominant emission component, while all the other Balmer lines are visible mostly in absorption.
Along with \ion{Ca}{ii} and \ion{O}{i}, we also identify a forest of metal lines, including the \ion{Fe}{ii} multiplets, \ion{Ba}{ii} ($\lambda\lambda$ 5853.7, 6141.7, 6496.9~\AA, \ion{Ti}{ii} ($\lambda\lambda$ 4160-4600~\AA), \ion{Sc}{ii} ($\lambda\lambda$ 5031.0, 5526.7-5684.2, 6245.6-6370.4~\AA) and a very prominent \ion{Na}{i}~D in absorption, now dominated by the higher-velocity component due to the expansion of the LRN ejecta. 
We note that none of these spectra show the [\ion{Ca}{ii}] doublet in emission, a feature typically observed in other types of gap transients, such as ILRTs \citep{Valerin_2024b}.

\subsection{Spectral evolution}

In this section, the evolution of the most prominent lines in the spectra of AT~2021aess is discussed. The complete sequence of low-resolution optical spectra obtained for AT~2021aess, corrected for redshift and reddening, is shown in Fig.~\ref{fig:spectra_evol}.

The spectral continuum, which is blue at the time of the first light-curve peak, becomes redder with time. The black-body temperature ($T_{\text{cont}}$) is computed through a black-body fit to the spectral continuum and, as expected, shows a similar evolution to that derived from the black-body fit to the SED (see Figs. \ref{fig:2021aess_T_L_R_colored} and \ref{fig:T_L_R_comparison.PDF}). The discrepancy in the two temperature estimates remains below $\leq 10\%$ during the first peak, and even less at later phases.

\begin{figure}
    \centering
    \includegraphics[width=1\linewidth]{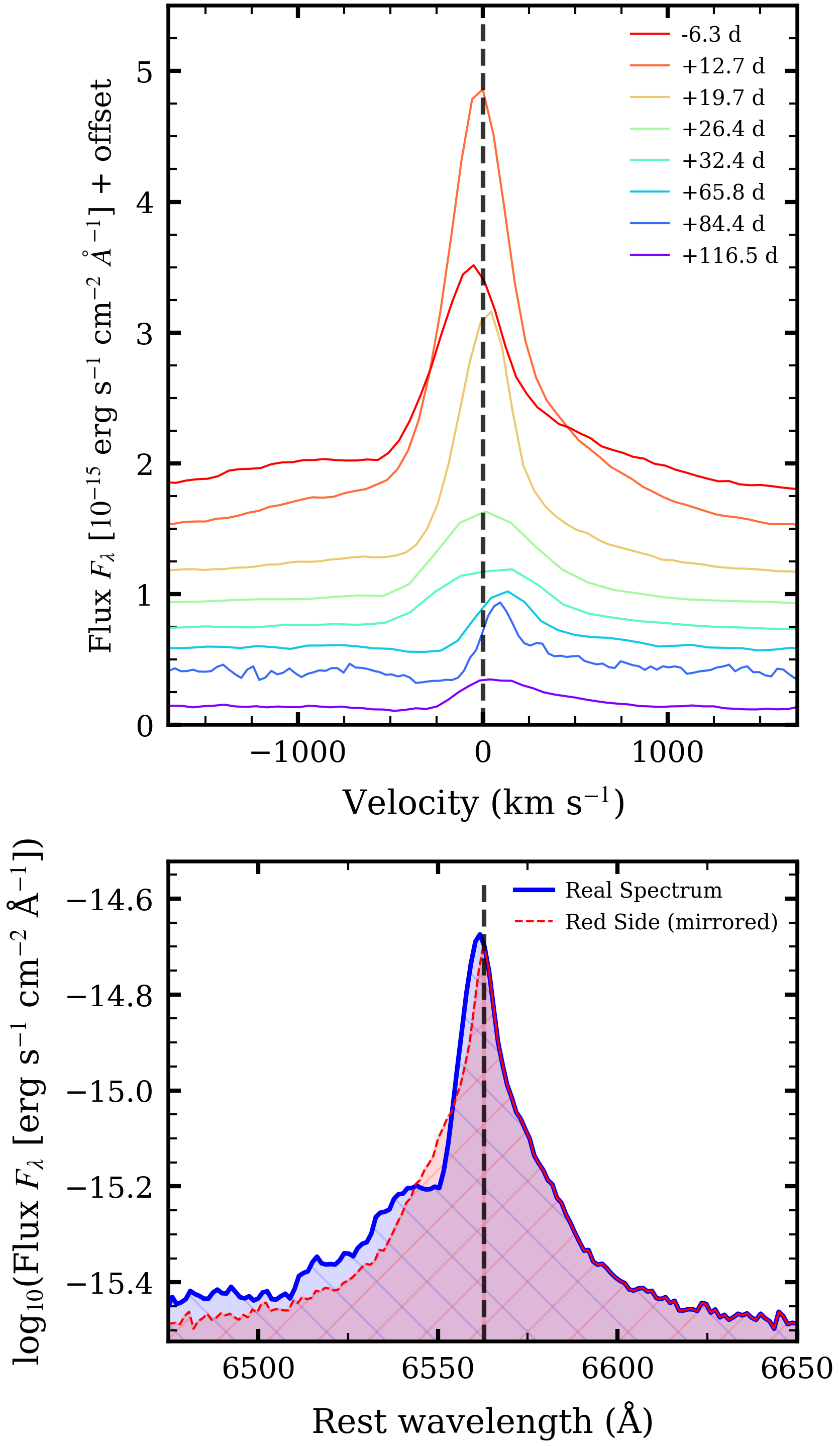}
    \caption{Top panel: Evolution of the H$\alpha$ line in the velocity space (only the spectra with high resolution are showed). The rest wavelength position is marked with a black dashed line. The flux initially increases when approaching the first peak and then decreases with time. Bottom panel: H$\alpha$ mirrored line of the $-6.3$~days spectrum, showing a profile with a noticeable asymmetry. The blue wing appears suppressed, likely due to an absorption component, possibly a P~Cygni profile, partially absorbing the blueshifted emission. This effect may be enhanced by the modest spectral resolution, which smooths and blends line features.
    \label{fig:H_line}}
\end{figure}

The spectra obtained from $-6.3$ to $+32.4$ days exhibit a temperature evolution from $T_{\text{cont}} \sim 9650\,\mathrm{K}$ shortly before maximum, down to $T_{\text{cont}} \sim 6500\,\mathrm{K}$ about one month after the outburst.
A narrow H$\alpha$ in emission is the most prominent feature in all spectra.
Signs of asymmetry are visible in its profile in the best-quality spectra. A close view of the evolution of the H$\alpha$ line in the velocity space is shown in Fig.~\ref{fig:H_line} (top panel).
H$\beta$ and other hydrogen lines are much weaker than H$\alpha$, and their emission components disappear about one month after the maximum.
The values of the fluxes of H$\alpha$ and H$\beta$, along with the FWHM velocity of H$\alpha$ (corrected for spectral resolution), are reported in Table \ref{tab:spectral_param}.
The lack of significant temporal evolution in the redward extent of the $\mathrm{H}\alpha$ profile (Table \ref{tab:spectral_param}) points toward a kinematically stable line-emitting region and ejecta geometry, where the extended red wing might be shaped by electron scattering within a dense envelope whose optical depth varies slowly over the observed epochs.

From the flux ratio of H$\alpha$ to H$\beta$, we compute the Balmer decrement at different epochs, up to $+32.4$ days, since this is the phase of the last available spectrum where H$\beta$ is clearly distinguishable. The evolution of the Balmer decrement can provide insights into the properties of the emitting gas. In the simplified scenario of Case~B recombination, with the gas being optically thick to photons of the Lyman series, the expected value for the Balmer decrement is $\approx 3$, with a weak dependence on the temperature of the recombining gas.
The Balmer decrement of AT~2021aess shows an increase from 1.7 at phase $-6.3$ days to about 3.0 at phase $+19.7$ days. Later, the Balmer decrement rises more rapidly up to 7.9 at phase $+32.4$ days. We note that the Balmer decrement values, around maximum luminosity, are close to the expectations from the Case B recombination. The increase in the Balmer decrement observed after the first peak is caused by the sharp decrease of the H$\beta$ flux. This could be interpreted as a sign of an increasing extinction along the line of sight, which would attenuate the flux of H$\beta$ in a stronger fashion than that of H$\alpha$. 
Alternatively, this steepening could be the result of high optical depth in the Balmer lines combined with collisional excitation. In dense media ($n_e \gtrsim 10^9\ \mathrm{cm^{-3}}$), large optical depths in Ly$\alpha$ and H$\alpha$ lead to a significant population of the $n=2$ level of neutral hydrogen. Collisional excitation from $n=2$ to $n=3$ requires less energy ($\Delta E = 1.89$~eV) than to $n=4$ ($\Delta E = 2.55$~eV) and is therefore strongly favoured at typical electron temperatures ($T_e \approx 10^4$~K), substantially boosting H$\alpha$ relative to H$\beta$. Furthermore, Balmer self-absorption and radiative transfer effects in an optically thick expanding envelope can suppress H$\beta$ more efficiently, driving the Balmer decrement to values well above the Case~B expectation (e.g., \citealt{Osterbrock,Drake_1980}).

On the other hand, the low value of the Balmer decrement measured in the first phases might be due to an overestimation of the interstellar reddening affecting the target, given that we are assuming dust properties similar to those of the Milky Way, or it could be explained with a relatively low optical depth of the Ly$\alpha$ line 
\citep[see page 96 and Fig.~4.3 of][]{Osterbrock}.
Multiple P~Cygni \ion{Fe}{ii} lines are detected in the spectral sequence of AT~2021aess, with the most prominent being those of the multiplet 42 triplet. Due to the fact that the spectra were obtained with different instruments and grisms, the observed widths of the spectral features are affected by different instrumental broadening. For each spectrum, we estimated the instrumental profile from the FWHM of the \ion{O}{i} $\lambda$5577 night-sky line (see Table~\ref{spectra_log}) and used this value to correct the measured line widths. In particular, intrinsic widths were derived after subtracting the instrumental resolution in quadrature. Features whose observed widths are comparable to the instrumental profile are unresolved or only marginally resolved; therefore, only velocity variations significantly larger than the instrumental resolution are considered physically meaningful.

In Table~\ref{tab:spectral_param}, we report the velocity computed from the minimum of the observed P~Cygni profile of H$\alpha$, H$\beta$ along with those inferred for the \ion{Fe}{ii} lines.
The P~Cygni velocity of H$\alpha$ is $390 \pm 40$~km~s$^{-1}$ at phase $-6.3$ days (see Fig. \ref{fig:H_line}, bottom panel), fading to $147\pm40$ km~s$^{-1}$ at +115.5 days. A similar temporal evolution is observed for the other measurable transitions. At a given epoch, the velocities inferred from H$\alpha$ and the \ion{Fe}{II} lines are generally compatible within the uncertainties, while H$\beta$ is often less well constrained. This behaviour is not inconsistent with a linear expansion
(i.e., at a given time, the velocity of the ejecta is proportional to their distance from the centre of the explosion), but suggests that the portion of the ejecta probed by these lines is characterised by a relatively shallow velocity gradient, so that different species trace regions with comparable expansion velocities.
As the photosphere recedes in mass coordinates, the material responsible for the absorption shows a progressively lower velocity. 

\begin{figure}
    \centering
    \includegraphics[width=1\linewidth]{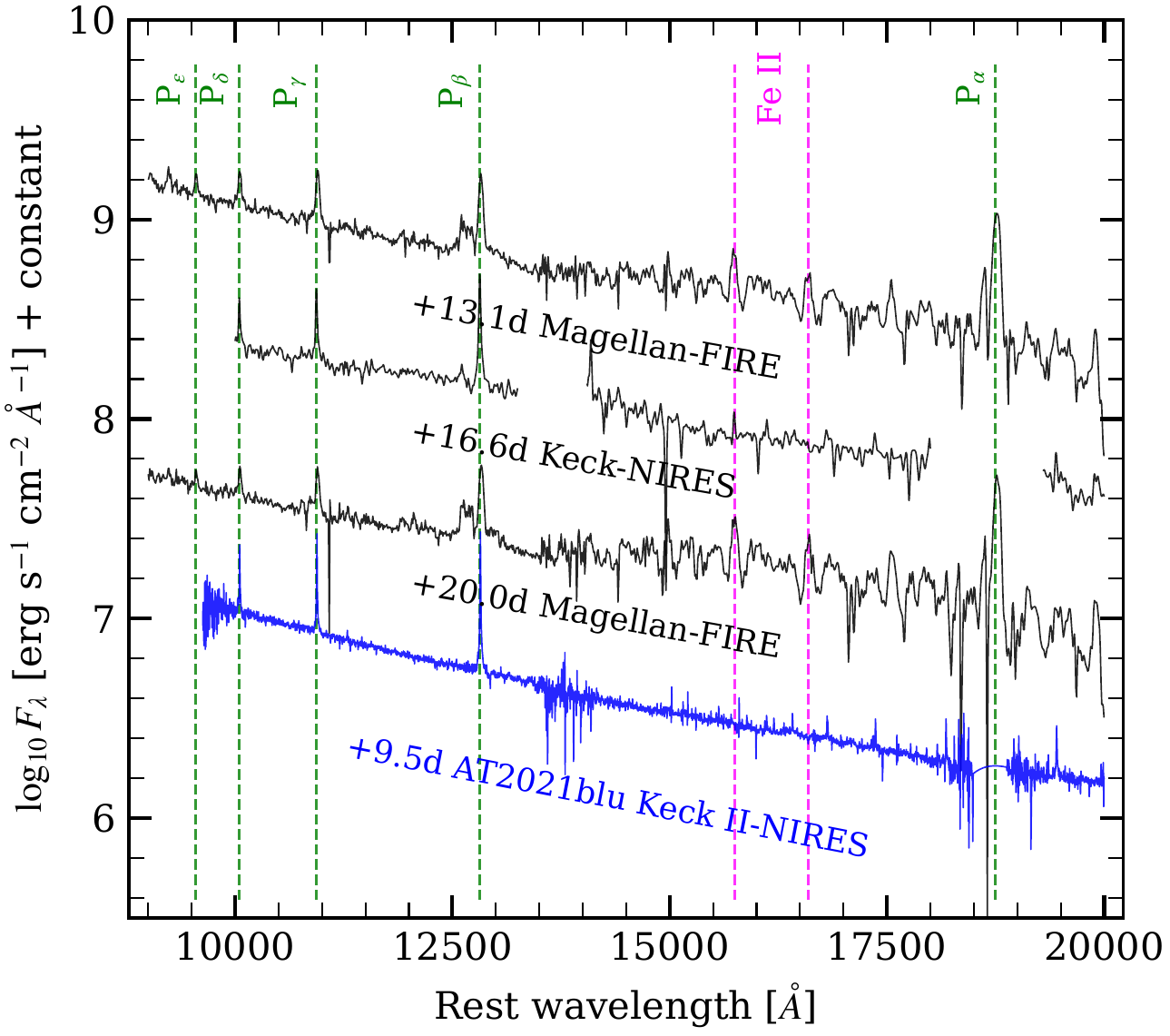}
    \caption{Spectral evolution in the NIR domain, with identification of the most prominent features. An almost coeval spectrum of LRN AT~2021blu  \citep{Pastorello_2023} is shown in blue as a comparison. }
    \label{fig:NIR_spectra_evol}
\end{figure}

In the earliest spectra, particularly at phases $-6.3$, $+12.7$, and $+19.7$ days, H$\alpha$ exhibits an asymmetry in its profile,with an evident blue shoulder.
The H$\alpha$ profile can hence be phenomenologically described by three components: a main emission with extended wings, a P-Cygni absorption and an additional broader blue-shifted emission. In the context of LRNe, the additional blue-shifted component could be related to a jet emitted during stellar coalescence. Alternatively, in an ejecta-CSM interaction scenario, an intermediate-width component may arise in shocked material. 
The attenuation of the red-shifted side of the profile may be due to prompt dust formation in the post-shock cool dense shell, as observed in some SNe IIn  \citep{Tartaglia_2020}.
 
This interpretation is consistent with the classical dust-formation signature proposed by \citealt{Lucy_1989}, where progressive extinction of receding material induces red-wing suppression and an apparent blue asymmetry of emission lines. Similar diagnostics have been discussed in interacting SNe (e.g. \citealt{Mattila_2008}; \citealt{Nozawa_2008}; \citealt{Smith_2008,Smith_2024}) and modeled in detail for eruptive transients using non-spherical ejecta geometries (e.g., \citealt{Shore_2018}). We stress, however, that our current Lorentzian+Gaussian deblending is a phenomenological description and does not constrain the large-scale geometry (symmetry, opening angle, or clump filling factor). Therefore, both an additional jet-like component and asymmetric ejecta/CSM interaction with dust attenuation  remain plausible explanations for the early H${\alpha}$ profile mismatches \citep{Stritzinger_2012}.
Although line-of-sight velocity gradients or local optical depth variations in a spherical shell cannot be excluded, such profile distortions could also tentatively reflect a non-spherical or multi-component ejecta geometry, similar to the structured outflows (HVR/LVR) modeled by \citealt{Mason_&_Shore_2022} for V1309 Sco.

At the time of the light-curve minimum between the two photometric peaks on +38.4 days, the spectra have become progressively redder and the continuum temperature $T_{\text{cont}}$ is $\sim$ 6500 K.
At this phase, the intensity of the H$\alpha$ line decreases, while H$\beta$ is no longer visible in emission.
After the first peak, the continuum temperature $T_{\text{cont}}$ exhibits a slow, linear decline from $\sim 6500$~K to $\sim 5000$~K, corresponding to phases $+32.4$ to $+114.5$~days. During this cooling phase, a ``forest'' of absorption lines of metals, including \ion{Ba}{ii}, \ion{Sc}{ii}, \ion{Ti}{ii} and \ion{Fe}{ii}, begins to dominate the spectra.

Although the spectral continuum reddens with time, the temperature remains always higher than that of other LRNe at similar phases (see Fig. \ref{fig:T_L_R_comparison.PDF}). This hotter continuum at late phases is reminiscent of that of SNhunt~248 \citep{Mauerhan, Kankare_2015}. However, this trend cannot be confirmed after the second peak, as no spectra are available in that phase. During the rapid photometric decline following the second peak, we expect the emergence of molecular bands and, possibly, an evolution toward a spectrum of an M-type star, as usually observed for other LRNe \citep[][and references therein]{Pastorello_2023}. However, despite the lack of late-time spectra, we are confident that  AT~2021aess evolves in a similar fashion as other LRNe, since photometric observations at very late phases clearly show the strong NIR excess for AT~2021aess observed in other objects of this class.

Three epochs of lower S/N NIR spectroscopy were obtained with the 6.5-m Magellan Baade telescope with the Folded-port InfraRed Echelette (FIRE) and the 10-m Keck II telescope with the Near-Infrared Echellette Spectrometer (NIRES). The last one was also presented by \cite{Medler2025ApJS..281...28M}.
Unfortunately, these spectra span a very short period of the LRN evolution, from +13.1 to +20.0~days from the maximum light. They are quite similar (see Fig.~\ref{fig:NIR_spectra_evol}), with an overall blue continuum with strong superposed Paschen lines, as well as evident \ion{Fe}{ii} features in the $H$-band region.

\section{Characterization of the progenitor system}\label{Characterization of the progenitor system}
\subsection{Analysis of archival images}\label{sect:archive_photometry}

In order to characterize the progenitor system of AT~2021aess, we searched for deep images available in the public data archives. The following data sets were identified: 
the 4\,m `Blanco' telescope at Cerro Tololo Inter-american Observatory (CTIO) equipped with the DECam instrument imaged the AT~2021aess site at several epochs between 2012 and 2017, during the DESI Legacy Imaging Surveys \citep{DESI2016arXiv161100036D, Dey2019AJ....157..168D}; the 2.65\,m VLT Survey Telescope (VST) equipped with the OmegaCAM wide-field camera observed NGC~1359 in November~2014; the 8.2\,m ESO-VLT (UT3) telescope equipped with the VIsible Multi-Object Spectrograph (VIMOS) imaged NGC~1359 in October 2007 and February 2008; the 3.58\,m Canada-France-Hawaii Telescope (CFHT) equipped with CFH12K CCD mosaic camera observed the LRN field in December 2002; NIR images of the LRN site were taken on July 14, 2016 with the 4.1\,m Visible and Infrared Survey Telescope for Astronomy (VISTA) equipped with the VISTA InfraRed CAMera (VIRCAM). 

We searched for signatures of a stellar source at the coordinates of AT~2021aess, and successfully detected a precursor in the 2012-2017 stacked DECam images, in the deep single-epoch 2007-2008 VLT images, in the 2002 CFHT image and in the 2014 OmegaCAM ($r$~band only) image. The NIR VIRCAM observations, instead, returned only shallow upper limits.
PSF-fitting photometry with \texttt{SNOoPY} was performed on the source at the LRN location. Detections and limits of the AT~2021aess precursor are reported in Table \ref{Stack}.

\subsection{Progenitor's analysis}\label{sect:progenitor_analysis}
The photometry of the LRN precursor described in Sect. \ref{sect:archive_photometry} reveals that the source at the location of AT 2021aess progressively increased its optical brightness from the earliest observation in 2002 to the latest one in 2012-2017. This implies that the progenitor of AT~2021aess cannot be considered quiescent in the 2007 to 2017 decade. As a consequence, the earliest CFHT image (2002 December 6) probably provides the best approximation of the quiescent stage for the LRN progenitor.

\begin{figure}[h]
    \centering
    \includegraphics[width=1\linewidth]{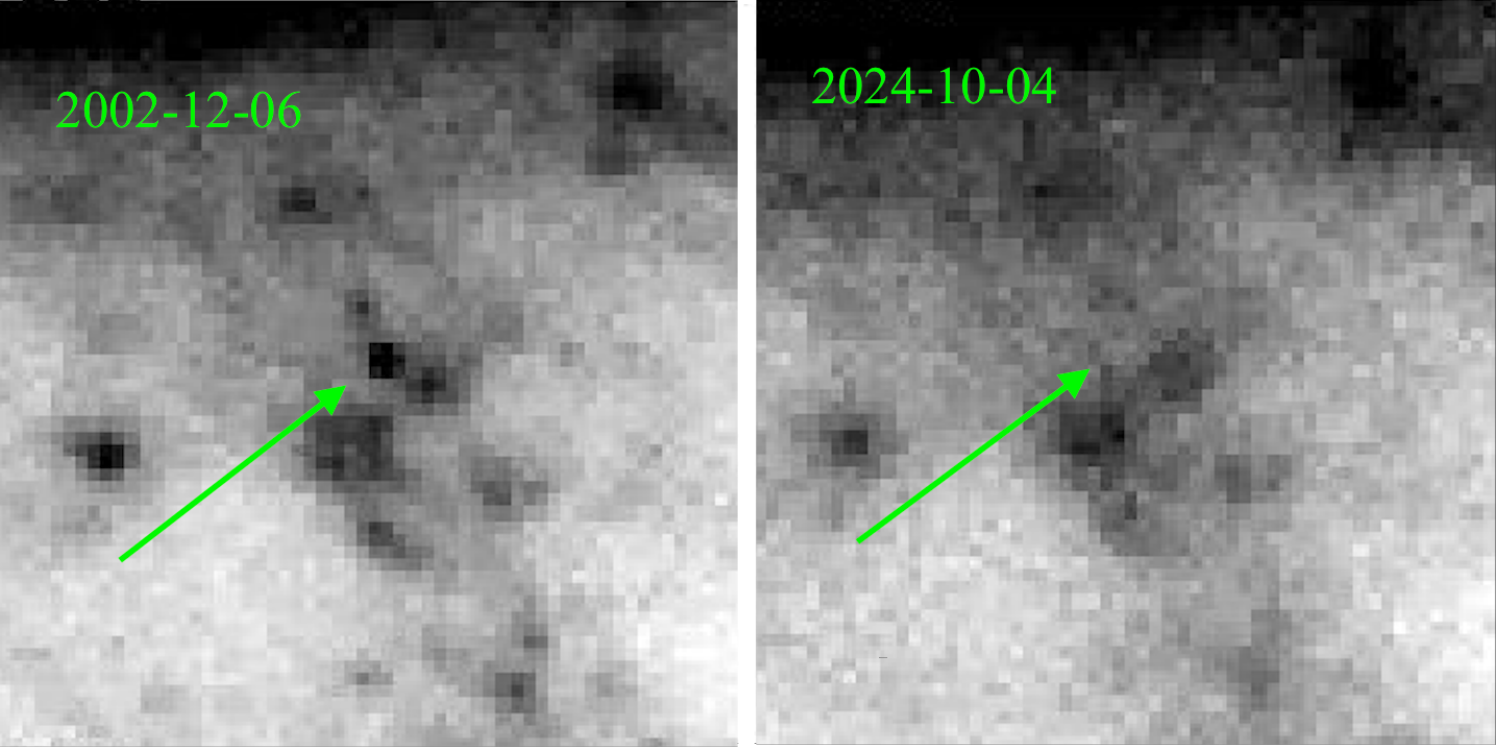}
    \caption{Comparison between red-band images of the transient site taken on 2002 December 6 ($R$ filter, with CFHT) and 2024 October 4 (Sloan $r$ filter, with NTT). A source, likely the progenitor, is clearly visible in 2002, whereas it is no longer detected in 2024.}
    \label{fig:progenitor_comparison}
\end{figure}

On 2024 October 4, approximately three years after the outburst, we obtained an $r$-band image with NTT+EFOSC2, with a total integration time of 18 minutes.
No source is detected at the location of the transient, returning an upper limit of $r > 23.1$ mag. A comparison between the images at the location of the progenitor on 2002 December 6 and 2024 October 4 is shown in Fig.~\ref{fig:progenitor_comparison}. The object detected in 2002 has disappeared, hence ruling out the source was due to background contamination. This unequivocally confirms that the source detected in 2002 and in the following years is the progenitor system of AT~2021aess. 

As LRNe are believed to be produced by violent stellar mergers, the outcoming star is expected to survive the core coalescence. The non-detection of a stellar source in the 2024 optical images is therefore somewhat unexpected. We note, however, that the $r$-band non-detection alone is not an unquestionable constraint on the remnant flux, because it may be influenced by the structured local background. The current optical non detection is thus consistent with the formation of a dusty circumstellar environment obscuring the merger remnant at optical wavelengths, a scenario that is further discussed in Sect.~\ref{Discussion} and can be tested through future IR observations.

To constrain the bolometric luminosity and the spectral type of the progenitor identified in pre-LRN images, we need to convert the reddening-corrected magnitudes into fluxes. This allows us to determine the SED of the source. To improve the SED sampling of the progenitor, we merge the photometric data obtained in two time intervals: from 2002 to 2008, and from 2012 to 2017. Given the slow photometric evolution of the progenitor candidate over time, we consider this approximation acceptable. The two SEDs are then fitted with different blackbodies, and this process returns approximate estimates of the color temperature and the bolometric luminosity, as reported in Table \ref{Stack_2}.

\begin{figure}
\centering
    \includegraphics[width=1.06\linewidth,angle=0]{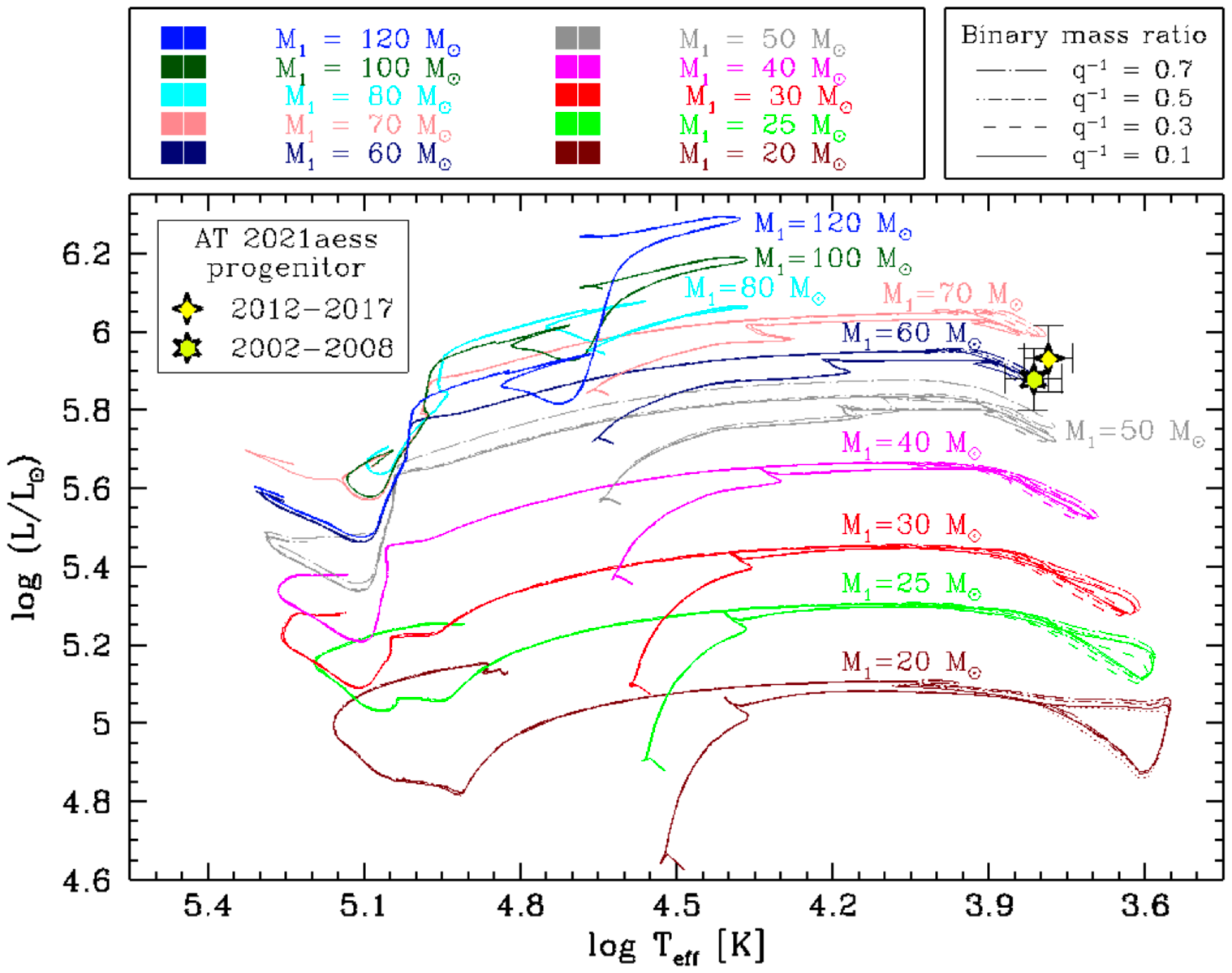}
    \caption{Location of the progenitor of AT~2021aess in the HRD, with yellow symbols with different shapes corresponding to observations at different epochs 
    and comparison with binary evolution tracks ($Z~=~0.02$), and orbital period $P \sim 250$ days.  }
    \label{fig:HRD}
\end{figure}

\begin{table}[]
  \centering
  \caption{Physical parameters of the progenitor obtained from black-body fits to the SED constrained from pre-outburst observations of the source at the position of AT~2021aess in 2002-2008 and 2012-2017. \label{Stack_2}}
  \vspace{0.3cm}
    \begin{tabular}{c|c|c|c}
    \toprule
    Date & \textit{T}~[K] & $L_{\text{bol}}$~[erg s$^{-1}$] & \textit{R}~[AU]\\
    \midrule
    2002-2008 & $6540 \pm 810$ & $2.92 (\pm 0.52) \times 10^{39}$ & $3.17 \pm 0.62$\\
    2012-2017 & $6110 \pm 650$ & $3.28 (\pm 0.63) \times 10^{39}$ & $3.84 \pm 0.69$\\
    \bottomrule
    \end{tabular}
    \vspace{0.3cm}
    \begin{flushleft}
    \end{flushleft}
\end{table}

Following \citet{Pastorello_2023}, we can obtain the mass of the candidate progenitor of AT~2021aess using the $T$ and $L_{\text{bol}}$ values reported in Table \ref{Stack_2} to place the source in a Hertzsprung-Russell diagram (HRD, see Fig. \ref{fig:HRD}). The location of the source in two periods (2002-2006 and 2012-2017) is then compared with a grid of Binary Population And Spectral Synthesis (BPASS, v2.2.1) evolutionary-track models at solar metallicity ($Z$~=~0.02) provided by \citet{Eldridge_2017} and \citet{Stanway_2018} {\footnote{The evolutionary tracks in the figure can be retrieved from \url{https://www.bpass.auckland.ac.nz/bpass-v2-2/}}.} The choice of these specific tracks is motivated by the approximately solar metallicity inferred for the environment of AT~2021aess (see Appendix \ref{Host galaxy metallicity estimation}), although we note that the progenitor constraints are more sensitive to the assumed binary configuration than to modest metallicity variations. Given the high intrinsic luminosity of the progenitor candidate, we considered only binary tracks for very massive primaries, with zero-age main-sequence (ZAMS) masses M$_{\text{ZAMS}} \geq 20\, $M$_\odot$.

To obtain an indicative constraint on the progenitor mass, Fig.~\ref{fig:HRD} shows a set of binary evolutionary tracks with primary masses in the range $20\, $M$_\odot \leq $ M$_{\text{ZAMS}} \leq 100\, $M$_\odot$, an initial orbital period of about 250~d, and four different binary mass ratios, corresponding to secondaries with masses between 10\% and 70\% of the primary mass. Among the parameters explored here, the initial orbital period has the strongest impact on whether the tracks intersect the observed progenitor locus in the HR diagram. In particular, models with shorter initial periods remain systematically too hot to reproduce the observed source, whereas larger periods would also allow for an intersection but with a substantially larger degeneracy. By contrast, the mass of the secondary has only a modest effect on the morphology of the track in the region of interest, and very few constraints can be posed on its mass\footnote{\citealt{Moe_&_Di_Stefano_2017} demonstrated that $f(q)$  strongly depends on the primary mass and the orbital period. For massive binaries with short-to-intermediate orbital periods ($P \lesssim 1000\text{ d}$) - which are the direct progenitors of contact binaries and common-envelope/merger events like LRNe - the distribution becomes progressively biased toward higher mass ratios, favoring companions of slightly lower or comparable masses ($0.3 < q < 1$).}.
Therefore, the comparison shown in Fig.~\ref{fig:HRD} should not be interpreted as a unique evolutionary solution, but rather as a representative estimate within the family of models explored here. Under these assumptions, the progenitor is consistent with a binary system whose primary had M$_{\text{ZAMS}} \approx 60 \pm 10\, $M$_\odot$.

To verify the robustness of the above mass estimate for the progenitor of AT~2021aess, we can use an empirical relation between the absolute magnitude of the LRN and the progenitor mass originally proposed by \citet{kochaneck_2014}, and later updated by \citet{Blagorodnova_2021}. According to the revised version proposed by \citet{Cai_2022a}\footnote{further updated by \cite{Reguitti_2026b}.} and based on the evidence of a robust correlation between the mass of the primary and the absolute $V$-band  magnitude of the LRN at the time of the second light-curve maximum, the mass of the primary can be inferred through the following relation:
\begin{equation} \label{eq1}
\mathrm{log_{10}}\,\biggl(\frac{\text{M}_{\rm prog}}{{\rm \text{M}}_\odot}\biggl) = (-0.172\pm0.017)\, \cdot \text{M}_V -(0.811\pm0.042).
\end{equation}
\noindent

The apparent $V$-band magnitude of AT~2021aess at the second light curve peak is $V = 18.44 \pm 0.01$ (see Table \ref{mag_peak_2nd}). Adopting the distance and total colour excess discussed in Sect.~\ref{Discovery and host galaxy}, an absolute magnitude at the second maximum of M$_V = -14.11 \pm 0.32$~mag is obtained.
From Eq. \ref{eq1}, we derive the following mass for the progenitor of AT~2021aess: M $= 41.3^{+32.1}_{-18.1} \, $M$_\odot$.
This mass estimate is significantly smaller, but still marginally consistent with that obtained through the comparison of the putative quiescent progenitor with BPASS evolutionary tracks (see above).
However, we caution that the progenitor mass inferred from the \citet{Cai_2022a} relation may be affected by significant uncertainty in the case of AT~2021aess, due to the unusually large luminosity contrast between the first and second peaks in its light curve. Such a peculiarity may suggest deviations from the typical LRN evolution used to calibrate that relation, potentially impacting the reliability of the derived mass estimate.

The progenitor system of AT2021aess (estimated at $\sim 50-65\,$M$_\odot$) belongs to the high-mass tail of progenitors that produce LRN outbursts, with a mass comparable to those of the most luminous events, such as SNhunt248 ($58\pm2\, M_\odot$), UGC12307-2013OT1 ($54^{+67}_{-30}\, M_\odot$), and AT~2021afy ($46^{+64}_{-25}\, M_\odot$)\citep[see the LRN mass estimates reported in Table~4 of][]{Pastorello_2023}.

\section{Discussion and concluding remarks}\label{Discussion}

LRNe are commonly interpreted as the outcome of binary interaction culminating in common-envelope ejection and stellar coalescence \citep{Tylenda_2011,Ivanova_2013c,Ivanova_2017,Molnar_2017}. Systems such as V1309~Sco provide a key empirical support for this framework, where unstable mass transfer and Roche-lobe overflow were observed prior to the outburst and interpreted as the immediate precursor to the merger event \citep{Mason_2010,Tylenda_2011,Mason_&_Shore_2022}. In this context, the first blue peak is often associated with the release of thermal energy from fast, low-mass ejecta and/or with interaction between the ejecta and pre-existing circumstellar material, while the later red peak or plateau is linked to the cooling and recombination of the material \citep{Tylenda_2006,Pejcha_2016,MacLeod_2017,MacLeod_2018,Matsumoto_2022}.

The previous sections presented the extensive optical and NIR spectro-photometric dataset of AT~2021aess, one of the best-sampled and most luminous ($L_{\text{bol}} = 7.4 \times 10^{41}$ erg~s$^{-1}$ at peak) LRNe observed to date. Overall, the observed behaviour of AT~2021aess is consistent with the general phenomenology of LRNe. Its light curve shows the characteristic double-peaked morphology of this class, with a first blue peak that is about four times more luminous than the second one. 
The photometry coverage is extensive and densely sampled, although monitoring of the photometric decline after the second maximum was interrupted due to the solar conjunction of the event.

The spectroscopic evolution is similar to that of other extra-galactic LRNe \citep{Pastorello_2019_review, Pastorello_2023} with early-time spectra dominated by a hard continuum ($T_{\text{cont}}$\,$\sim$\,$10^4$\,K) and prominent, narrow Balmer lines in emission. The spectrum evolves with time, becoming progressively redder and, at the epoch of the second light-curve maximum, it resembles that of a late G-type star. The Balmer lines are now much weaker, while several narrow metal lines appear, mostly in absorption. Again, this spectral evolution is similar to that normally observed in LRNe. However, in contrast with other objects \citep[e.g.,][]{Blagorodnova_2017}, the color temperature remains fairly high during the second light-curve peak ($\sim$5500~K, see Fig. \ref{fig:2021aess_T_L_R_colored}), and molecular bands of \ion{TiO}{} and \ion{VO}{} are not observed in our latest optical spectra, in analogy to another high-luminosity object in the literature \citep[SNhunt248;][]{Kankare_2015}.
However, we cannot rule out that the expected spectral transition towards a late-type (M-type like) star happened at later epochs, when the object was unobservable because of the seasonal gap.

The above speculation on the late-time spectral appearance is supported by the observations collected at phases $>$220 days, when the LRN field was observed again. AT~2021aess was no longer detectable in the optical bands, while it remained a luminous source in the NIR domain. This late photometric behaviour is not unusual \citep[see][for a discussion on the IR excess in evolved LRNe]{Blagorodnova_2021}, and it is usually interpreted as the consequence of dust formation. We remark that a dusty envelope embedding the merger candidate was observed, for instance, in the HST images of V838~Mon \citep{Bond_2003}.

As discussed in Sect. \ref{sect:archive_photometry}, the progenitor of AT~2021aess is likely to be at the more massive end of the distribution of LRNe progenitor systems (50-65 M$_\odot$) and underwent mass loss through binary interaction. This process led to the ejection of a common envelope, a progressive decrease in the angular momentum of the system, and ultimately to a merger event. This scenario is supported by the characteristic light-curve shape of the transient and its spectral evolution. The pre-coalescence mass loss is also supported by the detection of a faint, variable source at the LRN location, which showed an uninterrupted, slow brightening lasting almost two decades (see Fig.~\ref{fig:Absolute_magnitude_aess_main_spread_extended}, and the analysis of the photometric data of the progenitor over 15 years performed in Sect. \ref{sect:progenitor_analysis}). 
The AT~2021aess site was later imaged in the Sloan~$r$-band (October 4, 2024) and no source was visible at the LRN position down to $m_r = 23.1$ mag. This optical fading may suggest that the merger remnant has become obscured, possibly by newly formed dust. A broadly similar post-merger behaviour, with strong dust obscuration and a marked redistribution of the flux toward the IR, was observed in V1309 Sco \citep[e.g.,][]{kaminski_2018}, although the two transients are not necessarily direct analogues. 
We note, however, that the $r$-band non-detection alone is not a strong constraint on the remnant thermal emission and may also be influenced by the structured local background, including nearby H II-region emission.  Therefore, the current optical non-detection should be regarded only as being consistent with dust formation in the circumstellar environment \citep[see,][]{Mauerhan,Blagorodnova_2017,Blagorodnova_2020,Reguitti_2026}. This scenario should be verified through future IR observations of the AT~2021aess site, in particular with the James Webb Space Telescope (JWST; \citealt{Lau2025jwst.prop.7040L}).

LRNe are a heterogeneous class of transients that can be produced by progenitor systems spanning an enormous range of masses. So far, fewer than 30 extragalactic LRNe have been discovered, but only four of them have been observed in our Galaxy in the past $\sim$35 years. The Galactic LRNe are all intrinsically faint, hence associated with low-mass progenitors. \cite{kochaneck_2014} determined that low-luminosity events produced by low-mass progenitors are common, and LRNe similar to V1309~Sco are expected to occur in the Galaxy once every decade.
However, as pointed out by \cite{Karambelkar_2023}, the volumetric rate of LRNe has a strong dependence on their luminosity, and very luminous events are intrinsically rare. The strong correlation between the luminosity of the outburst and the progenitor mass system \citep[][]{Blagorodnova_2021, Cai_2022a} implies that LRNe originating from massive systems are rare. Following the predictions of \citet[][see their Figure 12]{Karambelkar_2023}, we estimate that LRNe with luminosity comparable to that of AT~2021aess have a rate of the order of $10^{-7}$ Mpc$^{-3}$ year$^{-1}$, 4-5 orders of magnitude less frequent than faint events such as V1309~Sco. 
The advent of JWST and the Euclid spacecraft, along with facilities such as the Vera Rubin Observatory and the Nancy Grace Roman Space Telescope, will significantly improve LRN rate estimates and provide unprecedented insights into the nature of LRNe and the fate of their progenitors.

\section*{Data availability}

The spectra presented in this work are publicly available through the WISeREP repository \citep{Ofer2012} and can be downloaded from \url{https://www.wiserep.org/object/19838}. 
Photometry presented in this work and provided in Tables \ref{mag_peak_1st} and \ref{Stack} are available in electronic form at the CDS via anonymous ftp to cdsarc.u-strasbg.fr (130.79.128.5) or via \url{http://cdsweb.u-strasbg.fr/cgi-bin/qcat?J/A+A/}.

%Table B2 is only available at the CDS via anonymous ftp to cdsarc.cds.unistra.fr (130.79.128.5) or via https://cdsarc.cds.unistra.fr/viz-bin/cat/J/A+A/XXX/AXXX}

\begin{acknowledgements}

We thank the referee for his thorough and constructive reports, which significantly helped to improve the quality of this work. We thank the staff of the various observatories where data were obtained for their assistance.
FG, GV, AR, YZC, IS acknowledge financial support from the SOXS project (PI S. Campana).

AR, AP, GV, NE-R, I.S. acknowledge financial support from the PRIN-INAF 2022 ``Shedding light on the nature of gap transients: from the observations to the models''. 

This work made use of v2.2.1 of the Binary Population and Spectral Synthesis (BPASS) models as last described in \cite{Stanway_2018}.

Based in part on observations collected at the European Organisation for Astronomical Research in the Southern Hemisphere, Chile, under ESO programs 080.B-0361 (VIMOS), 177.A-3011 (OMEGACam), 179.A-2010 (VIRCam), and as part of ePESSTO+ (the advanced Public ESO Spectroscopic Survey for Transient Objects Survey - PI: Inserra), under ESO programs ID 108.220C and 112.25JQ.

The Pan-STARRS1 Surveys (PS1) and the PS1 public science archive have been made possible through contributions by the Institute for Astronomy, the University of Hawaii, the Pan-STARRS Project Office, the Max-Planck Society and its participating institutes, the Max Planck Institute for Astronomy, Heidelberg and the Max Planck Institute for Extraterrestrial Physics, Garching, The Johns Hopkins University, Durham University, the University of Edinburgh, the Queen's University Belfast, the Harvard-Smithsonian Center for Astrophysics, the Las Cumbres Observatory Global Telescope Network Incorporated, the National Central University of Taiwan, the Space Telescope Science Institute, the National Aeronautics and Space Administration under Grant No. NNX08AR22G issued through the Planetary Science Division of the NASA Science Mission Directorate, the National Science Foundation Grant No. AST-1238877, the University of Maryland, Eotvos Lorand University (ELTE), the Los Alamos National Laboratory, and the Gordon and Betty Moore Foundation.
Pan-STARRS is a project of the Institute for Astronomy of the University of Hawaii, and is supported by the NASA SSO Near Earth Observation Program under grants 80NSSC18K0971, NNX14AM74G, NNX12AR65G, NNX13AQ47G, NNX08AR22G, 80NSSC21K1572 and by the State of Hawaii.

ZTF is supported by the National Science Foundation under Grants No. AST-1440341 and AST-2034437 and a collaboration including current partners Caltech, IPAC, the Oskar Klein Center at Stockholm University, the University of Maryland, University of California, Berkeley , the University of Wisconsin at Milwaukee, University of Warwick, Ruhr University, Cornell University, Northwestern University and Drexel University. Operations are conducted by COO, IPAC, and UW.

MD Stritzinger is funded by the Independent Research Fund Denmark (IRFD, grant number 10.46540/2032-00022B) and by an Aarhus University Research Foundation Nova project (AUFF-E-2023-9-28).

T-WC acknowledges the financial support from the Yushan Fellow Program by the Ministry of Education, Taiwan (MOE-111-YSFMS-0008-001-P1) and the National Science and Technology Council, Taiwan (NSTC grant 114-2112-M-008-021-MY3).

Y-ZC is supported by the National Natural Science Foundation of China (NSFC, Grant No. 12303054), the Yunnan Fundamental Research Projects (Grant Nos. 202401AU070063, 202501AS070078), the National Key Research and Development Program of China (Grant No. 2024YFA1611603), and the International Centre of Supernovae (ICESUN), Yunnan Key Laboratory of Supernova Research (No. 202505AV340004)

LG acknowledges financial support from AGAUR, CSIC, MCIN and AEI 10.13039/501100011033 under projects PID2023-151307NB-I00, PIE 20215AT016, CEX2020-001058-M, ILINK23001, COOPB2304, and 2021-SGR-01270.\\

JIGH acknowledges financial support from the Spanish Ministry of Science, Innovation and Universities (MICIU) projects PID2020-117493GB-I00 and PID2023-149982NB-I00.\\
This material is based upon work supported by the National Science Foundation Graduate Research Fellowship Program under Grant Nos. 1842402 and 2236415. Any opinions, findings, conclusions, or recommendations expressed in this material are those of the author(s) and do not necessarily reflect the views of the National Science Foundation.\\

CPG acknowledges financial support from the Secretary of Universities and Research (Government of Catalonia) and by the Horizon 2020 Research and Innovation Programme of the European Union under the Marie Sk\l{}odowska-Curie and the Beatriu de Pin\'os 2021 BP 00168 programme, from the Spanish Ministerio de Ciencia e Innovaci\'on (MCIN) and the Agencia Estatal de Investigaci\'on (AEI) 10.13039/501100011033 under the PID2023-151307NB-I00 SNNEXT project, from Centro Superior de Investigaciones Cient\'ificas (CSIC) under the PIE project 20215AT016 and the program Unidad de Excelencia Mar\'ia de Maeztu CEX2020-001058-M, and from the Departament de Recerca i Universitats de la Generalitat de Catalunya through the 2021-SGR-01270 grant.

Based on observations made with the Nordic Optical Telescope, owned in collaboration by the University of Turku and Aarhus University, and operated jointly by Aarhus University, the University of Turku and the University of Oslo, representing Denmark, Finland and Norway, the University of Iceland and Stockholm University at the Observatorio del Roque de los Muchachos, La Palma, Spain, of the Instituto de Astrofisica de Canarias. The NOT data were obtained under program IDs P64-507, P66-506.

Partly based on observations collected at the Copernico 1.82m Telescope operated by INAF - Osservatorio Astronomico di Padova at Cima Ekar, Asiago, Italy.

\end{acknowledgements}

  \bibliographystyle{aa} 
  \bibliography{bibliography}

\begin{appendix}

\section{Host galaxy metallicity estimation}\label{Host galaxy metallicity estimation}
\begin{figure}
    \centering
    \includegraphics[width=1\linewidth]{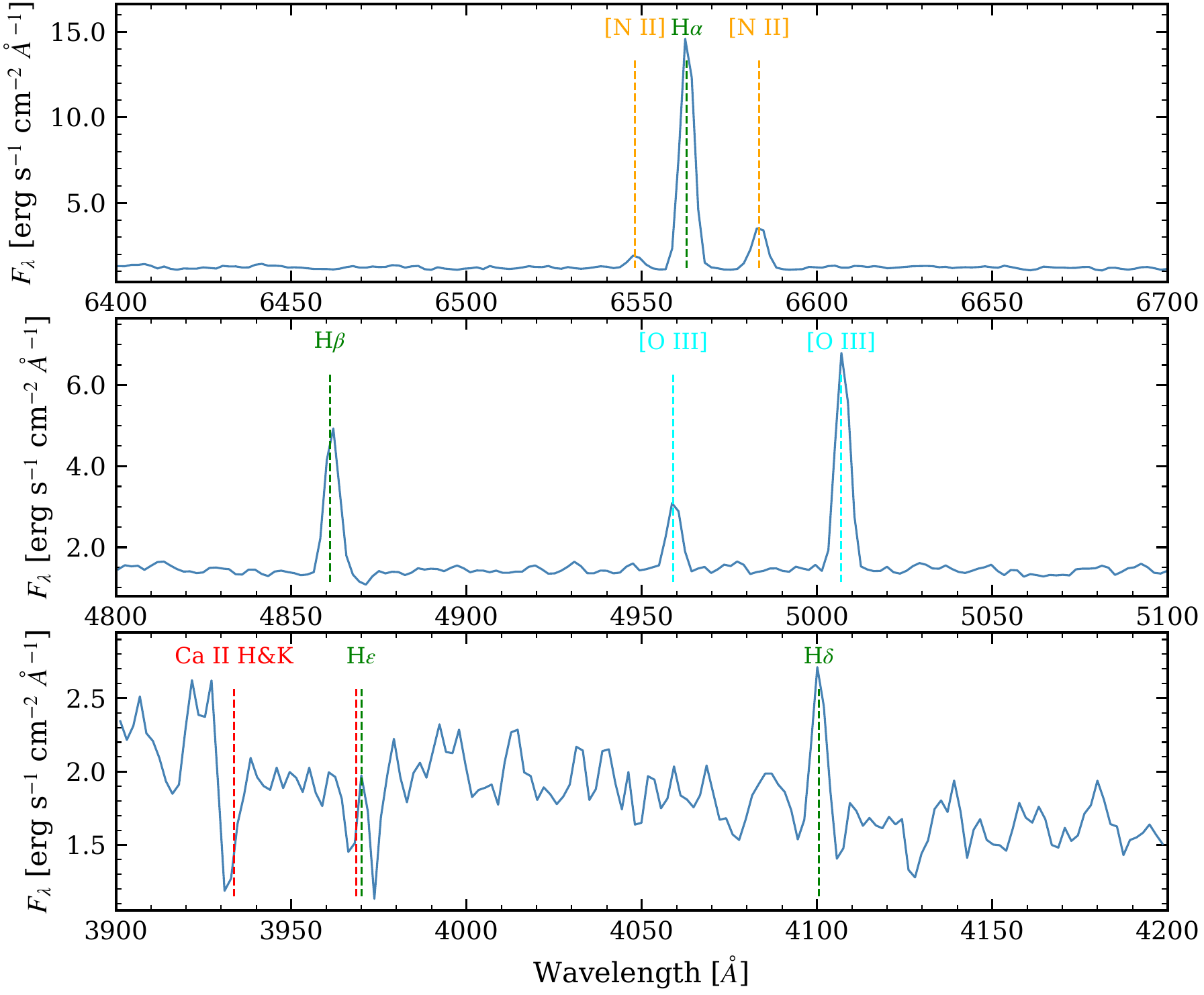}
    \caption{Spectrum of the host galaxy NGC 1359 taken at 294.5 days after the $V$-band LRN light-curve maximum. The fluxes are multiplied by a factor $10^{-14}$. }
    \label{fig:galaxy}
\end{figure}

The INT+IDS spectrum taken on September 16, 2022 shows no evident features attributable to AT~2021aess (see Fig.\,\ref{fig:galaxy}). However, we note the presence of narrow, unresolved emission lines in the proximity of the expected LRN site. These lines are associated with an \ion{H}{ii} region serendipitously included in the slit, and can be used to derive an approximate estimate of the oxygen abundance in the local environment sampled by the slit.

Several strong-line diagnostics may be used to derive an approximate oxygen abundance for this \ion{H}{ii} region. Following \citet{Pettini_2004}, we first define the N2 index as the logarithm of the intensity ratio between \ion{N}{ii} $\lambda$\,6583 and H$_\alpha$, that is,

\begin{equation} \label{eq_met1}
N2 = \text{log}\,\biggl(\frac{I(\ion{N}{ii}_{\lambda6583})}{I(\text{H}_\alpha)}\biggl)~,
\end{equation}
\noindent

\noindent for which we obtain N2\,$\approx$\,$-0.7$. Adopting the calibration reported by \citet{Denicolo_2002}, the local oxygen abundance is given by 
\begin{equation} \label{eq_met2}
12 + \text{log}\,\biggl(\frac{\ion{O}{}}{\ion{H}{}}\biggl) = 9.12 + 0.73 \times N2 = 8.61~.
\end{equation}
\noindent

We also estimated the oxygen abundance through the O3N2 method \citep{Pettini_2004}:

\begin{equation} \label{eq_met3}
O3N2 = \text{log}\,\biggl(\frac{I(\ion{O}{iii}_{\lambda5007})/I(\text{H}_\beta)}{I(\ion{N}{ii}_{\lambda6583})/I(\text{H}_\alpha)}\biggl)~.
\end{equation}
\noindent

\noindent After measuring the intensities of the lines necessary to determine the O3N2 index, following \citet{Pettini_2004}, the oxygen abundance is provided by the following relation: 
\begin{equation} \label{eq_met4}
12 + \text{log}\,\biggl(\frac{\ion{O}{}}{\ion{H}{}}\biggl) = 8.73 - 0.32 \times O3N2 = 8.56~,
\end{equation}
\noindent

The two diagnostics provide mutually consistent values, indicating an oxygen abundance of about $12 + \text{log}(O/H) \approx 8.6$. Given the uncertainties inherent in strong-line calibrations, and considering that the spectrum samples a finite \ion{H}{ii} region rather than the exact transient position, we conservatively conclude that the local environment is approximately consistent with solar metallicity within the uncertainties \citep{Asplund2021A&A...653A.141A}.

\begin{figure}[!h]
    \centering
    \includegraphics[width=1\linewidth]{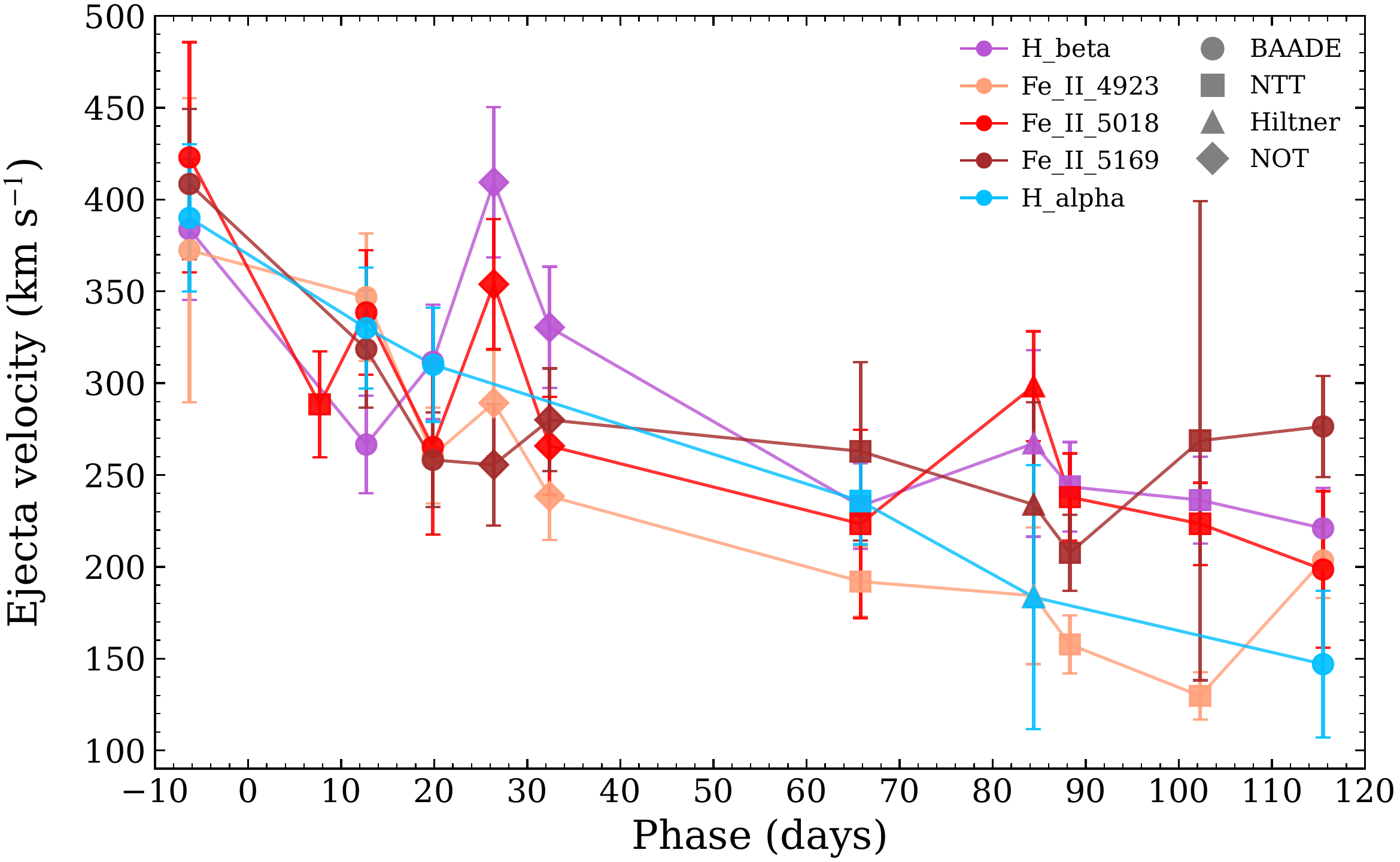}
    \caption{Time evolution of the P~Cygni derived velocity for $H_\alpha$, $H_\beta$ and the three \ion{Fe}{ii} multiplet~42 $\lambda\lambda$\,4924,5018,5169 lines.}
    \label{fig:P_Cygni_vel}
\end{figure}

\newpage
\onecolumn
\section{Tables} 
\label{appendix}
\begin{table*}[h]
  \centering
  \caption{Log of the spectroscopic observations. The phases are relative to the $V$-band light-curve peak.\label{spectra_log}}
 \vspace{0.3cm}
    \begin{tabular}{c|c|c|c|c|c|c}
    \toprule
    Date & MJD & Phase & Instruments  & Range & Grism & Res.$^\star$\\
    
    [yyyy-mm-dd] & [J2000] & [days] & & [$\AA$] & & [$\AA$] \\

    \addlinespace[1pt]
    \midrule
    2021-11-20 & 59538.21 & $-6.3$ & Baade+IMACS & 4200 - 9400 & Gri-300 & 5.0 \\
    2021-12-04 & 59552.19 & 7.7 & NTT+EFOSC2 & 3600 - 9200 & gr13 & 17.5 \\
    2021-12-09 & 59557.27 & 12.7 & Baade+IMACS & 4200 - 9400 & Gri-300 & 5.0 \\
    2021-12-09 & 59557.59 & 13.1 & Baade+FIRE & 7800 - 25500 & LDPrism & $\cdots$ \\
    2021-12-13 & 59561.19 & 16.6 & Keck II+NIRES & 9300 - 25000 & 110 & $\cdots$  \\
    2021-12-16 & 59564.22 & 19.7 & Baade+IMACS & 4200 - 9400 & Gri-300 & 5.0 \\
    2021-12-16 & 59564.59 & 20.0 & Baade+FIRE & 7800 - 25500 & LDPrism & $\cdots$  \\
    2021-12-23 & 59570.95 & 26.4 & NOT+ALFOSC & 3500 - 9800 & gr4 & 14.8 \\
    2021-12-28 & 59576.90 & 32.4 & NOT+ALFOSC & 3800 - 9800 & gr4 & 13.8 \\
    2022-01-03 & 59582.90 & 38.4 & NOT+ALFOSC & 3800 - 9800 & gr4 & 18.4 \\
    2022-01-10 & 59589.87 & 45.3 & NOT+ALFOSC & 3000 - 8999 & gr4 & 18.0 \\
    2022-01-22 & 59602.13 & 57.6 & NTT+EFOSC2 & 3000 - 9999 & gr16 & 13.8 \\
    2022-01-31 & 59610.15 & 65.8 & NTT+EFOSC2 & 3800 - 9800 & gr18 & 7.6 \\
    2022-02-02 & 59613.08 & 69.3 & NTT+EFOSC2 & 3800 - 9800 & gr13 & 27.0 \\
    2022-02-09 & 59620.09 & 76.3 & NTT+EFOSC2 & 3800 - 9800 & gr13 & 17.6 \\
    2022-02-19 & 59629.59 & 84.4 & MDM+OSMOS\footnote{Hiltner 2.4-meter Telescope} & 4200 - 6800 & blue & 2.8 \\
    2022-02-22 & 59633.36 & 88.3 & NTT+EFOSC2 & 3800 - 9800 & gr13 & 18.5 \\
    2022-03-09 & 59647.01 & 102.3 & NTT+EFOSC2 & 3800 - 9800 & gr13 & 17.6 \\
    2022-03-21 & 59659.00 & 114.5 & NTT+EFOSC2 & 3800 - 9800 & gr13 & 17.5 \\
    2022-03-22 & 59660.00 & 115.5 & Baade+IMACS & 4200 - 9400 & Gri-300 & 5.0 \\
    2022-09-16 & 59839.15 & 294.5 & INT+IDS & 3800 - 8300 & R300V & 2.4   \\
    \bottomrule
    \end{tabular}
    \vspace{0.1cm}
    \vspace{0.1cm}
    \begin{flushleft}
    \tablefoot{
        $^\star$ The quoted values correspond to the FWHM of the  [O~I] $\lambda$5577 night-sky line, adopted as an empirical estimate of the instrumental profile (line-spread function) for each spectrum.\\
        TELESCOPES: NTT = New Technology Telescope, 3.58\,m (ESO, La Silla, Chile); 
        NOT = Nordic Optical Telescope, 2.56\,m (Roque de Los Muchachos, La Palma, Spain); 
        Baade = Magellan Baade Telescope, 6.5\,m (Las Campanas Observatory, Chile); 
        Keck II = W. M. Keck Observatory, 10.0\,m (Mauna Kea, Hawaii, USA); 
        INT = Isaac Newton Telescope, 2.54\,m (Roque de Los Muchachos, La Palma, Spain); 
        MDM = Hiltner Telescope, 2.4\,m (MDM Observatory, Kitt Peak, Arizona, USA). \\
        INSTRUMENTS: IMACS = Inamori-Magellan Areal Camera \& Spectrograph; 
        FIRE = Folded-port InfraRed Echellette; 
        NIRES = Near-Infrared Echellette Spectrometer; 
        EFOSC2 = ESO Faint Object Spectrograph and Camera v.2; 
        ALFOSC = Alhambra Faint Object Spectrograph and Camera; 
        IDS = Intermediate Dispersion Spectrograph;
        OSMOS = Ohio State Multi-Object Spectrograph.}
    \end{flushleft}
    
\end{table*}

\onecolumn

\begin{table}
  \centering
  \caption{Telescopes and instruments used for the photometric follow-up campaign of AT~2021aess. \label{Telescopes}}
  \vspace{0.3in}
    \begin{tabular}{c|c|c|c|c}
    \toprule
    Telescope & Site & Instruments & Scale ($\arcsec$/px) & Filters \\
    \addlinespace[0pt]
    \midrule
    3.58~m NTT & European Southern Observatory (La Silla, Chile) & SOFI & 0.288 & JHKs \\
    2.56~m NOT & Roque de Los Muchachos (La Palma, Spain) & NOTCam & 0.234 & JHKs \\
    3.5~m CAHA & Calar Alto (Almer\'ia, Spain) & OMEGA2000 & 0.450 & H \\
    10.0~m Keck II & Mauna Kea (Hawaii, USA) & NIRES & 0.135 & K \\
    4.1~m VISTA & Paranal Observatory (Chile) & VIRCAM & 0.340 & JK \\
    5.1~m Hale (P200) & Palomar Observatory (California, USA) & TSpec & ---- & JHK$^{\dagger}$ \\
    11.0~m SALT & South African Astronomical Observatory (Sutherland, South Africa) & HRS & ---- & JHK$^{\dagger}$ \\
    \midrule
    %\hline
    3.58~m NTT & European Southern Observatory (La Silla, Chile)  & EFOSC2 & 0.240 & V \\
    2.00~m LT & Roque de Los Muchachos Obs. (La Palma, Spain) & IO:O & 0.150 & BVugriz \\
    2.56~m NOT & Roque de Los Muchachos Obs. (La Palma, Spain) & ALFOSC & 0.213 & BVgriz \\
    10.4~m GTC & Roque de Los Muchachos Obs. (La Palma, Spain)  & OSIRIS & 0.254 & ugriz \\
    2.54~m INT & Roque de Los Muchachos Obs. (La Palma, Spain)  & WFC & 0.330 & BVgriz \\
    1.82~m CT  & Asiago Observatory (Mt. Ekar, Italy) & AFOSC & 0.260 & iz \\
    1.0~m ST & Las Campanas Observatory (Chile) & E2V & 0.435 & BVugri \\
    4.02~m VMBT & Cerro Tololo Inter-American Observatory (Chile) & DECam & 0.263 & griz \\
    3.58~m CFHT & Mauna Kea Observatory (Hawaii Islands, USA) & CFH12K & 0.206 & R \\
    8.20~m VLT-UT3 & European Southern Observatory (Cerro Paranal, Chile) & VIMOS & 0.205 & BVI \\
    \midrule
    \rowcolor{white}
    \multicolumn{5}{c}{Wide-field Surveys} \\
    \midrule
    1.8~m PS1 & Haleakalā Observatory (Hawaii, USA) & GPC1 & 0.258 & iwy \\
    1.2~m ZTF & Palomar Observatory (California, USA) & ZTF Camera & 1.010 & gri \\
    \bottomrule
    \end{tabular}
   \tablefoot{
   NTT = New Technology Telescope; NOT = Nordic Optical Telescope; CAHA = Centro Astron\'omico Hispano-Alem\'an;
   LT = Liverpool Telescope; GTC = Gran Telescopio Canarias; INT = Isaac Newton Telescope; CT = Copernico Telescope;
   ST = Henrietta Swope Telescope; VMBT = V\'ictor Manuel Blanco Telescope; CFHT = Canada-France-Hawaii Telescope;
   VLT-UT3 = Very Large Telescope (Melipal UT3); VISTA = Visible and Infrared Survey Telescope for Astronomy; 
   PS1 = Panoramic Survey Telescope and Rapid Response System 1; ZTF = Zwicky Transient Facility;
   Keck = W.M. Keck Observatory; P200 = Hale Telescope.; SALT = Southern African Large Telescope;\\
   $^{\dagger}$Photometric magnitudes derived from spectrophotometry of NIR spectra published by \citet{Karambelkar_2023}.
   }
\end{table}

\begin{table}[h]
  \centering
  \caption{Magnitude rise/decline rates from the light-curve peaks in the different bands, in mag (100 days)$^{-1}$, ordered by effective wavelength.\label{rates}}
    \begin{tabular}{c|c|c|c|c}
    \toprule
    Filter & rise 1 & decline 1 & rise 2 & decline 2 \\
    \addlinespace[0pt]
    \midrule
    $u$ & $-$23.10 $\pm$ 1.75 & 12.38 $\pm$ 0.36 & - & - \\
    $B$ & $-$20.97 $\pm$ 1.73 & 8.11 $\pm$ 0.23 & $-$0.24 $\pm$ 0.06 & 2.96 $\pm$ 0.16 \\
    $g$ & $-$23.00 $\pm$ 1.33 & 7.35 $\pm$ 0.24 & $-$0.88 $\pm$ 0.07 & 3.12 $\pm$ 0.17 \\
    $V$ & $-$20.72 $\pm$ 1.53 & 7.51 $\pm$ 0.21 & $-$0.71 $\pm$ 0.06 & 2.51 $\pm$ 0.14 \\
    cyan & - & -& $-$0.80 $\pm$ 0.15 & - \\
    $r$ & $-$21.25 $\pm$ 1.53 & 7.96 $\pm$ 0.17 & $-$0.85 $\pm$ 0.03 & 2.24 $\pm$ 0.11 \\
    $i$ & $-$23.45 $\pm$ 1.35 & 7.95 $\pm$ 0.16 & $-$1.03 $\pm$ 0.03 & 1.92 $\pm$ 0.14 \\
    orange & $-$20.00 $\pm$ 2.71 & 7.12 $\pm$ 0.53 & $-$1.11 $\pm$ 0.13 & 1.33 $\pm$ 0.38 \\
    $z$ & - & - & $-$1.09 $\pm$ 0.46 & - \\
    $J$ & - & - & $-$0.63 $\pm$ 0.07 & - \\
    $H$ & - & - & $-$0.90 $\pm$ 0.11 & - \\
    $K$ & - & - & $-$0.70 $\pm$ 0.09 & - \\
    \bottomrule
    \end{tabular}
    \vspace{0.3cm}
    \begin{flushleft}
    \end{flushleft}
\end{table}

\begin{table*}[h]
  \centering
  \caption{Photometry of the AT~2021aess progenitor candidate in pre-outburst archival images.\label{Stack}}
  \vspace{0.3cm}
  \begin{tabular}{c|c|c|c|c}
    \toprule
    Date & MJD$^\star$ & Band & $m$ & Instrument \\
    \addlinespace[0pt]
    \midrule
    2002-12-06 & 52614.29 & $R$ & $22.27 \pm 0.12$ & CFH12K \\
    2007-10-19 & 54392 & $I$ & $21.53 \pm 0.11$ & VIMOS \\
    2007-10-19 & 54392 & $V$ & $22.28 \pm 0.16$ & VIMOS \\
    2008-02-10 & 54506 & $B$ & $23.73 \pm 0.33$ & VIMOS \\
    2013-01-30 & 56322.09 & $z$ & $>$20.9 & OMEGACam \\
    2013-02-12 & 56335.06 & $i$ & $>$20.7 & OMEGACam \\
    2014-11-27 & 56988.22 & $r$ & $21.90 \pm 0.12$ & OMEGACam \\
    2014-12-25 & 57016.17 & $g$ & $>$22.9 & OMEGACam \\
    (2012-2017) & $57157 \pm 923$ & $g$ & $23.13 \pm 0.28$ & DECam \\
    (2012-2017) & $57157 \pm 923$ & $r$ & $21.99 \pm 0.09$ & DECam \\
    (2012-2017) & $57157 \pm 923$ & $i$ & $22.19 \pm 0.11$ & DECam \\
    (2012-2017) & $57157 \pm 923$ & $z$ & $21.99 \pm 0.11$ & DECam \\
    2016-07-14 & 57583.38 & $J$ & $>$20.5 & VIRCAM \\
    2016-07-14 & 57583.37 & $K$ & $>$18.4 & VIRCAM \\
    \bottomrule
  \end{tabular}

  \vspace{0.1cm}
  \begin{flushleft}
  \tablefoot{
    $^\star$ An average MJD is reported for stacked images. At the adopted host-galaxy distance of 26.06 Mpc, 1 arcsec corresponds to 126.3 pc. The nominal angular sampling of the instruments used here corresponds to projected physical scales of about 26.0 pc pix$^{-1}$ (CFH12K), 25.9 pc pix$^{-1}$ (VIMOS), 27.0 pc pix$^{-1}$ (OMEGACam), 33.2 pc pix$^{-1}$ (DECam), and 42.9 pc pix$^{-1}$ (VIRCAM). \\
    CFH12K = 12k$\times$8k Mosaic Camera on the 3.58~m Canada-France-Hawaii Telescope (CFHT; Mauna Kea Observatory, Hawaii, USA); \\
    VIMOS = Visible Multi-Object Spectrograph on the 8.2~m VLT-UT3 (Very Large Telescope, Melipal UT3; ESO, Paranal Observatory, Chile); \\
    OMEGACam = OmegaCAM on the 2.65~m VLT Survey Telescope (VST; ESO, Paranal Observatory, Chile); \\
    DECam = Dark Energy Camera on the 4.0~m V\'ictor M. Blanco Telescope (CTIO, Chile); \\
    VIRCAM = VISTA InfraRed CAMera on the 4.1~m VISTA Telescope (ESO, Cerro Paranal, Chile).
    }
  \end{flushleft}
\end{table*}

\begin{table}[t]
  \centering
  \caption{Evolution of the $H_{\alpha}$ and $H_{\beta}$ fluxes, the FWHM velocity of $H_{\alpha}$, and the velocities inferred from the positions of the Fe II (multiplet 42) P\,Cygni minima. \label{tab:spectral_param}}
  \begin{tabular}{c|cc|cccccc}
  \toprule
  Phase & F(H$\alpha$) & F(H$\beta$) & v$_{\mathrm{FWHM}}$(H$\alpha$) & H$\beta$ & Fe\,{\sc ii} $\lambda$4923 & Fe\,{\sc ii} $\lambda$5018 & Fe\,{\sc ii} $\lambda$5169 & H$\alpha$ \\
  
  [days] & [$10^{-15}$ erg s$^{-1}$] & [$10^{-15}$ erg s$^{-1}$] & [km s$^{-1}$] & [km s$^{-1}$] & [km s$^{-1}$] & [km s$^{-1}$] & [km s$^{-1}$] & [km s$^{-1}$] \\
  \midrule
  $-6.3$  & $44.0 \pm 4.0$  & $20.0 \pm 2.0$ & $660 \pm 60$ & $384 \pm 38$ & $372 \pm 83$ & $423 \pm 63$ & $408 \pm 41$ & $390 \pm 40$ \\
  +7.7   & $88.0 \pm 8.0$  & $44.0 \pm 4.0$ & $\leq 930$ & --- & --- & $288 \pm 29$ & --- & --- \\
  +12.7  & $62.0 \pm 6.0$  & $19.0 \pm 2.0$ & $450 \pm 45$ & $267 \pm 27$ & $347 \pm 35$ & $338 \pm 34$ & $319 \pm 32$ & $330 \pm 33$ \\
  +19.7  & $27.0 \pm 3.0$  & $8.4 \pm 0.8$ & $310 \pm 30$ & $312 \pm 31$ & $261 \pm 26$ & $265 \pm 48$ & $258 \pm 26$ & $310 \pm 31$ \\
  +26.4  & $13.0 \pm 1.0$  & $2.3 \pm 0.2$ & $\leq 800$ & $409 \pm 41$ & $289 \pm 29$ & $354 \pm 35$ & $256 \pm 33$ & --- \\
  +32.4  & $10.0 \pm 1.0$  & $1.3 \pm 0.1$ & $\leq 740$ & $330 \pm 33$ & $238 \pm 24$ & $266 \pm 27$ & $280 \pm 28$ & --- \\
  +38.4  & $7.0 \pm 0.7$   & --- & $\leq 1000$ & --- & --- & --- & --- & --- \\
  +45.3  & $6.1 \pm 0.6$   & --- & $\leq 960$ & --- & --- & --- & --- & --- \\
  +57.6  & $5.8 \pm 0.6$   & --- & $\leq 740$ & --- & --- & --- & --- & --- \\
  +65.8  & $6.6 \pm 0.6$   & $0.9 \pm 0.1$ & $\leq 350$ & $233 \pm 23$ & $192 \pm 19$ & $223 \pm 51$ & $263 \pm 49$ & $236 \pm 24$ \\
  +69.3  & $5.6 \pm 0.6$   & --- & $\leq 1450$ & --- & --- & --- & --- & --- \\
  +76.3  & $3.3 \pm 0.3$   & --- & $\leq 950$ & --- & --- & --- & --- & --- \\
  +84.4  & $6.7 \pm 0.7$   & --- & $330 \pm 30$ & $267 \pm 51$ & $184 \pm 37$ & $298 \pm 30$ & $234 \pm 56$ & $184 \pm 72$ \\
  +88.3  & $4.7 \pm 0.5$   & --- & $\leq 990$ & $244 \pm 24$ & $158 \pm 16$ & $238 \pm 24$ & $208 \pm 21$ & --- \\
  +102.3 & $5.0 \pm 0.5$   & --- & $\leq 950$ & $236 \pm 24$ & $130 \pm 13$ & $223 \pm 22$ & $269 \pm 130$ & --- \\
  +114.5 & $5.4 \pm 0.5$   & --- & $\leq 940$ & --- & --- & --- & --- & --- \\
  +115.5 & $6.7 \pm 0.7$   & --- & $440 \pm 40$ & $221 \pm 22$ & $204 \pm 20$ & $198 \pm 43$ & $276 \pm 28$ & $147 \pm 40$ \\
  \bottomrule
  \end{tabular}
  \vspace{0.3cm}

\end{table}

\end{appendix}

\end{document}